\documentclass{jaa}
\usepackage{natbib}
\usepackage{bibspacing}
\usepackage{graphicx}
\usepackage{xcolor}
\usepackage{upgreek}
\usepackage{hyperref}
\usepackage{soul}
\usepackage[normalem]{ulem}

\usepackage[pagewise, modulo]{lineno}

\newcommand\aj{AJ}

\newcommand\mnras{MNRAS}

\newcommand{\sigmarm}{\sigma_{\rm RM}}

\newcommand{\mkG}{{\rm \upmu G}}

\newcommand{\kpc}{\, {\rm kpc}}
\newcommand{\radm}{\,{\rm rad\,m^{-2}}}

\newcommand{\ghz}{\,{\rm GHz}}

\def\Pm{\rm Pm}

\begin{document}\sloppy

\title{\centering Exploring diffuse radio emission in galaxy clusters and groups with \\the uGMRT and the
SKA}


\author{Surajit Paul\textsuperscript{1,*}, Ruta Kale\textsuperscript{2}, Abhirup Datta\textsuperscript{3}, Aritra Basu\textsuperscript{4}, Sharanya Sur\textsuperscript{5}, Viral Parekh\textsuperscript{6}, Prateek Gupta\textsuperscript{1}, Swarna Chatterjee\textsuperscript{3}, Sameer Salunkhe\textsuperscript{1}, Asif Iqbal\textsuperscript{7}, Mamta Pandey-Pommier\textsuperscript{8}, Ramij Raja\textsuperscript{3}, Majidul Rahaman\textsuperscript{9,3}, Somak Raychaudhury\textsuperscript{10}, Biman B. Nath\textsuperscript{11}, and Subhabrata Majumdar\textsuperscript{12}
}
\affilOne{\textsuperscript{1}Department of Physics, Savitribai Phule Pune University, Pune 411007, India.\\}
\affilTwo{\textsuperscript{2}National Centre for Radio Astrophysics (NCRA), Tata Institute of Fundamental Research (TIFR), Pune 411 007, India.\\}
\affilThree{\textsuperscript{3}Department of Astronomy, Astrophysics and Space Engineering, Indian Institute of Technology Indore, Simrol, 453552, India.\\}
\affilFour{\textsuperscript{4}Thüringer Landessternwarte, Sternwarte 5,
07778 Tautenburg, Germany.\\}
\affilFive{\textsuperscript{5}Indian Institute of Astrophysics,
II Block, Koramangala,
Bengaluru 560 034, INDIA.\\}
\affilSix{\textsuperscript{6}Department of Physics and Electronics, Rhodes University, PO Box 94, Makhanda, 6140, South Africa.\\}

\affilSeven{\textsuperscript{7}AIM, CEA, CNRS, Universit\'e Paris-Saclay, Universit\'e Paris Diderot, Sorbonne Paris Cit\'e, F-91191 Gif-sur-Yvette, France\\}

\affilEight{\textsuperscript{8}Laboratoire Univers et Particules de Montpellier (LUPM)/CNRS,
Université de Montpellier, CC 072 - Place Eugène Bataillon
34095 Montpellier Cedex 5, France.\\}

\affilNine{\textsuperscript{9}Institute of Astronomy and Department of Physics, National Tsing Hua University, No. 101, Section 2, Kuang-Fu Road, Hsinchu 30013, Taiwan.\\}

\affilTen{\textsuperscript{10}Inter-University Centre for Astronomy and Astrophysics, Ganeshkhind, Pune 411007, India\\}

\affilEleven{\textsuperscript{11}Raman Research Institute, Raman Research Institute, Bangalore, 560080, India \\}
\affilTwelve{\textsuperscript{12}Tata Institute of Fundamental Research, 1 Homi Bhabha Road, Colaba, Mumbai 400005, India\\}


\twocolumn[{

\maketitle

\corres{surajit@physics.unipune.ac.in}

\msinfo{1 January 2022}{1 January 2022}

\begin{abstract}
Diffuse radio emission has been detected in a considerable number of galaxy clusters and groups, revealing the presence of pervasive cosmic magnetic fields, and of relativistic particles in the large scale structure of the Universe. Since the radio emission in galaxy systems is faint and its spectrum is steep, its observations are largely limited by the instrument sensitivity and frequency of observation, leading to a dearth of information, more so for lower-mass systems. The recent commissioning or upgrade of several large radio telescope arrays, particularly at the low frequency bands ($<$~GHz), is, therefore, a significant step forward. The unprecedented sensitivity of these new instruments, aided by the development of advanced calibration and imaging techniques, have helped in achieving unparalleled image quality and revolutionised the study of cluster-scale radio emission. At the same time, the development of state-of-the-art numerical simulations and the availability of supercomputing facilities have paved the way for high-resolution numerical modeling of radio emission, and the structure of the cosmic magnetic fields, associated with large-scale structures in the Universe, leading to predictions matching the capabilities of observational facilities.
In view of these rapidly-evolving developments in modeling and observations, in this review, we summarise the role of the new telescope arrays and the development of advanced imaging techniques and discuss the range of detections of various kinds of cluster radio sources, both in dedicated surveys as well as in numerous individual studies. We pay specific attention to the kinds of diffuse radio structures that have been able to reveal the underlying physics in recent observations. 
In particular, we discuss observations of large-scale sections of the cosmic web in the form of supercluster filaments, and studies of emission in low mass systems such as poor clusters and groups of galaxies, and of ultra-steep spectrum sources, the last two being notably aided by low-frequency observations and high sensitivity of the instruments being developed.
We also discuss and review the current theoretical understanding of various diffuse radio sources in clusters and the associated magnetic field and polarization in view of the current observations and simulations. As the statistics of detections improve along with our theoretical understanding, we update the source classification schemes based on the intrinsic properties of these sources. We conclude by summarising the role of the upgraded GMRT (uGMRT) and our expectations from the upcoming Square Kilometre Array (SKA) observatories.

\end{abstract}


}]


\doinum{12.3456/s78910-011-012-3}
\artcitid{\#\#\#\#}
\volnum{000}
\year{2022}
\pgrange{1--}
\setcounter{page}{1}
\lp{1}

\section{Radio observations of galaxy clusters}

Clusters of galaxies are massive ($\gtrsim 10^{14}\rm{M_{\odot}}$) and large ($\sim$ Mpc scale)  systems residing at the top of the hierarchy of extragalactic gravitationally-bound structures in the Universe. They are usually found embedded at the cross-roads of the cosmic web \citep{Weygaert_2008LNP,Springel_2006Natur} and many are found non-relaxed, indicating that they are still in the process of formation  \citep{Molnar_2016FrASS}. Various physical processes in the components of these systems make them luminous enough to be detected in astronomical observations at various wavelengths. Clusters are generally thermalized, their hot intra-cluster baryonic medium being prominent in X-ray emission \citep{Sarazin_1986RevModPhys}. A number of them are also detected as diffuse sources at radio wavelengths (for a review: \citealt{Weeren_2019SSRv}), confirming their non-thermal energy content. 

These diffuse continuum radio sources in galaxy clusters are associated with the intra-cluster medium (ICM) and are usually not associated with any optical counterparts in the system. Typically, their extent is in the range 100~kpc to a few Mpc with diverse morphologies. The emission usually has a power-law energy distribution with a steep spectral index\footnote{The spectral index, $\alpha$ is defined as $S_\nu\propto\nu^{-\alpha}$ 
where $S_\nu$ is the flux density at the frequency $\nu$.} ($\alpha
\!\sim\! 1.3$). These sources are generally faint at GHz frequencies 
($\sim$0.1--1 $\mu$Jy arcsec$^{-2}$ at 1.4 GHz), and the steepness of the spectrum is largely due to the nature of the injection spectra \citep{Stroe_2014MNRAS}. The evolution of the plasma associated with the radio emission is controlled largely by synchrotron and inverse Compton energy losses, and thus, as the plasma gets older, it results in further steepening of the radio spectrum \citep[e.g.][]{Weeren_2009A&A}. 

The cluster radio sources have been traditionally classified as radio halos, relics and mini-halos, depending mainly on their shape, the process of origin and the location within the cluster \citep[e. g.][]{Ensslin_1998A&A,Giovannini_1999NewA,Feretti_2008LNP,Feretti_2012A&ARv}. 
An updated version appears in a review by \citet{Weeren_2019SSRv}, where a new class has been added combining the revived AGN fossil plasma sources (``phoenix") and Gently re-energized tails (GReETs, \citealt{deGasperin_2017ScienceAdvances}), with a discussion on the characteristic properties of these sources. 

Since these sources are low in brightness and have a steep spectrum, their detection is limited by the instrument sensitivity and the sophistication of imaging techniques. Therefore, the study of diffuse radio emission from clusters got a huge fillip with the advent of highly sensitive telescopes as well as with the development of advanced calibration and imaging software and pipelines.

The review is organized as follows. The role of large telescopes is described in \S\ref{sec:telescopes-surveys} and the current software packages and methods are described in \S\ref{sec:packages_pipelines}. We further summarize the large surveys conducted so far and their future scope in \S\ref{sec:surveys}. The observational status of various kinds of cluster diffuse radio sources and the detection of associated magnetic fields have been updated in \S\ref{sec:cluster_radio_sources}. Recently, systematic studies of the detection of diffuse radio sources from structures apart from and beyond the clusters (such as super-cluster filaments, galaxy groups and poor clusters) have gained momentum. We summarize and update on these efforts in \S\ref{sec:groups_n_low-mass} and \S\ref{sec:bridge-supercluster}. Updates on the theoretical understanding of diffuse synchrotron radio emission, the role of AGN feedback and the origin and evolution of cosmic magnetic fields from various cluster sources along with the efforts made so far on numerical simulations on this matter have been presented in \S\ref{sec:theoretical_models_simulations}. With a better understanding of the origin of these radio sources, we propose a novel classification in \S\ref{sec:taxonomy}. The role of uGMRT and the expectations from the upcoming SKA telescopes in the context of clusters have been summarized in \S\ref{sec:role_uGMRT_SKA}, and conclusions outlined in \S\ref{sec:summary}.


\subsection{Role of large telescope arrays}\label{sec:telescopes-surveys}

The first large diffuse radio source was discovered in the nearby Coma cluster \citep{Large_1959Natur} with a size of 1.1 Mpc. Yet, to gather a sizeable sample of galaxy cluster scale diffuse radio emissions, it took more than five decades, mostly through targeted individual objects and a handful of dedicated surveys \citep{Venturi_2007A&A,Venturi_2008A&A,Venturi_2013A&A,Giacintucci2011,Kale_2013A&A,Kale_2015A&A,Parekh_2017MNRAS,Paul_2019MNRAS}. 

In recent times, however, there has been a race for searching diffuse radio emission in clusters, especially from large surveys. A large sample of 115 clusters, though many among them are previously known, were observed under The MeerKAT Galaxy Cluster Legacy Survey (MGCLS).  MeerKAT has allowed us to study them at an unprecedented depth of $3\mu$~Jy~beam$^{-1}$ at wide L-Band (900-1670 MHz) with full polarization information \citep{Knowles_2022A&A}, revealing many new unique features of diffuse structures. LoFAR Two Meter Sky Survey (LoTSS-1~\&~2), on the other hand, has given an opportunity to study the sky at low frequencies (central frequency of 144 MHz) with a median rms as low as 83~$\mu$Jy~beam$^{-1}$ \citep{Shimwell_2019A&A,Shimwell_2022A&A}. For deep fields, maps are further deeper at 30~$\mu$Jy~beam$^{-1}$ \citep{Osinga_2021A&A}. LoFAR has so far discovered more than $\sim60$ new clusters with halo, relics or revived fossil plasma sources (or candidates) \citep{Weeren_2021A&A,Osinga_2021A&A,Mandal_2020A&A,Botteon_2022A&A}. \citet{Botteon_2020MNRAS,Hoang_2019A&A} even discovered ultra-steep spectrum and extremely faint radio bridges between the pairs of interacting clusters. Studies with LoFAR helped in mapping diffuse radio emissions from the Supercluster members \citep{Ghirardini_2021A&A} as well as constraining magnetic field in the filaments \citep{Locatelli_2021A&A}. Strikingly, ultra-steep spectrum halos have been discovered in massive relaxed clusters \citep{Savini_2019A&A}. The ultra-sensitive wideband frequency coverage by uGMRT made possible the study of extremely steep ($\alpha>2.0$ and curved spectrum relic/phoenix emissions \citep{Kale_2018MNRAS,Lal_2020ApJS}. Low-mass clusters and groups that remained undetected due to their expected low luminosity have now been detected using the wideband uGMRT studies \citep{Paul_2021MNRAS} as well as in LoFAR, the emission from the intra-group medium \citep{Nikiel_2019A&A}. Spectral index studies of distant radio halos are done by combining data from uGMRT and LoFAR, two extremely sensitive telescopes working at complementary frequencies, \citep{Gennaro_2021A&A}.

\subsection{Role of imaging packages and pipelines}\label{sec:packages_pipelines}

A vital aspect of radio interferometry is processing the raw data from telescope observation and converting it to science-ready images. Owing to the low surface brightness of the cluster diffuse radio sources, dynamic range and sensitivity limitations at low frequencies makes them challenging targets. New generation radio interferometry comes with wideband wide-field observations, which have helped scan the sky over a large field of view with improved sensitivity. Due to the large bandwidth, the observing radio frequency bands are also often crowded with several undesirable radio frequency interference (RFI) which should be removed properly for an improved image. AOFlagger \citep{Offringa_2010MNRAS,Offringa_2012MNRAS} is an automated pipeline for effective RFI excision with a faster execution time. Moreover, the resolution, sensitivity to extended emission, and primary beam changes over the large bandwidth need to be accounted for in wideband observations. Wide-field imaging suffers non-coplanar baseline aberration, which results in degradation of image and phase error and thus needs to be corrected for the w-term. The introduction of wide-field, Multi-scale (MS) and Multi Term (MT) Multi Frequency Synthesis (MFS) imaging \citep{2011A&A...532A..71R} helped alleviate issues with wideband imaging and construct better spatial and spectral structure across a large field of view. The Astronomical Image processing System (AIPS), developed by NRAO (National Radio Astronomy Observatory), is one of the oldest imaging packages. While Common Astronomy Software Application (CASA: \citealt{McMullin_2007ASPC}), developed by NRAO in the last decade, is more user-friendly. With flagging, direction-independent calibration, imaging techniques, and post-deconvolution primary beam correction, CASA has been successful in providing science-ready images. The imaging algorithm of CASA includes MS-MFS, and MT-MFS techniques, as well as the W projection method \citep{2008ISTSP...2..647C} for correcting the effects from non-coplanar baselines. The joint channel deconvolution in WSClean Imager \citep{offringa_2014MNRAS,offringa_2017MNRAS} with w-stacking algorithm performs imaging 2-3 orders faster than CASA MSMFS. In addition to these, radio interferometric observations are hampered by a variety of time-dependent instrumental and ionospheric effects, particularly at low-frequency observations, and require direction-dependent calibration methods. These effects are well represented by Jones matrices and are now being addressed, thanks to upgraded calibration techniques. Source peeling and atmosphere modelling (SPAM: \citealt{Intema_2009A&A,Intema_2017A&A}) is a widely used pipeline for both direction independent and dependent calibration for GMRT. KillMS \citep{Tasse_2014ARXIV} and DD-FAcet \citep{Tasse_2018A&A}, Facet-based imaging software, are exceptional in dealing with externally defined direction-dependent Jones matrices and varying beam patterns for LoFAR. KillMS DD-FAcet can also be used for other telescopes like GMRT, VLA, MeerKAT, and ATCA. With the A-projection algorithm \citep{Bhatnagar_2008A&A}, the pointing errors and errors induced due to asymmetric antenna power pattern can be removed during deconvolution. 

Larger arrays increase the data size, escalating the computing load for imaging and calibration. Manual data reduction for these big data sets is both labour-intensive and error-prone, and thus requires modern processing pipelines. With the introduction of automated and semi-automated pipelines, for instance VLA Calibration pipeline for updated VLA data, SPAM and CAPTURE for GMRT (\citealt{Kale_2021ExperimentalAstronomy}), Apercal for WSRT data reduction \citep{Schulz_2020ASCL}, CARACal \citep{jozsa_2020ASPC} for MeerKAT, data handling has been made easier. The WSClean$+$IDG algorithm applies the correction in the image plane instead of the visibility plane and maps large sky areas with better efficiency and less computational cost \citep{Sweijen_2022NatAs}. Cubical \citep{Kenyon_2018MNRAS} is a cython-based calibration package designed for multiprocessing and much faster execution. Recent studies by \citealt{Wilber2020PASA,Parekh_2021Galaxies} have shown how the use of these software packages has significantly improved image dynamic range.

These highly sensitive telescopes and the development of advanced and sophisticated imaging packages and pipelines have helped in revealing a number of sources with new and unclassifiable properties, especially, plenty of ultra-steep and curved spectrum sources.
Also, the discovery of many low-mass clusters with diffuse radio sources, with widely differing values of radio power, exceeded theoretical expectations. Interestingly, though it took six decades of radio observations to discover about 200 diffuse radio sources associated with galaxy clusters, only the last three years of observations have revealed more than 100 new sources \citep{Botteon_2022A&A} with these new instruments, from even a limited sky coverage. LoFAR, from the Planck clusters list alone, is predicted to discover roughly 350 new clusters \citep{Botteon_2022A&A} with diffuse radio sources in the near future, from the entire northern sky.

\subsection{uGMRT and the SKA}\label{sec:uGMRT_SKA}

The GMRT is a radio telescope array located in Khodad, near Pune, India \citep{swarup_1991CSci}. It consists of 30 dish antennas, with a diameter of 45 m each, spread in a roughly ``Y"-shaped array forming baselines with lengths in the range of 100 m to 25 km. 
The upgrade of the GMRT involved the replacement of the earlier narrowband receivers with broadband ones and a new correlator (GMRT Wideband Backend) that allows recording data with instantaneous bandwidths of up to 400 MHz \citep{2017CSci..113..707G}. The wideband observations (400 MHz) as compared to the earlier narrowband ones (33 MHz) imply an improvement in sensitivity by a factor 3.4. Most importantly, in the context of imaging diffuse radio sources, it significantly improves the $uv$-coverage. A simulation study of imaging extended sources with the uGMRT shows that there will be considerable improvement in recovering the morphology of extended sources due to the improved $uv$-coverage \citep{2017ExA....44..165D}.

The SKA is proposed to be the largest radio observatory to be built. Phase 1 will consist of two parts: SKA1-mid and SKA1-low. The SKA1-mid will operate in the frequency range of 350 MHz--15 GHz and will consist of 197 dish antennas offering maximum baseline lengths of up to 150~km. The SKA1-low will operate at 50--350 MHz and will be an aperture array of 512 stations spread over an array offering maximum baselines of 65 km\citep{braun99}. The SKA will be able to image the sky south of declination $+44^{\circ}$. 

In the context of diffuse radio emission from galaxy clusters, the SKA 1 is expected to uncover radio halo emission in thousands of clusters up to the redshift of 0.6 \citep{2015aska.confE..73C,2016JApA...37...31K} and similarly for mini-halos \citep{2018A&A...617A..11G}, allowing to compare the status of magnetic fields in low and high redshift clusters. With SKA pathfinders such as the uGMRT and precursor facilities, we expect to get a glimpse of the possible science.

\section{Radio surveys of galaxy clusters}\label{sec:surveys}

The search for diffuse emission in galaxy clusters has been carried out systematically by either using all-sky radio surveys or targeted observations. 
The GMRT Radio Halo Survey (GRHS) carried out at 610 MHz was the first, large, targeted sub-GHz survey of galaxy clusters \citep{2007A&A...463..937V,2008A&A...484..327V}. The GRHS and its extension, together referred to as the EGRHS, consisted of 64 clusters selected from the REFLEX and eBCS catalogues 
with the criteria: X-ray luminosity $>5\times10^{44}$ erg s$^{-1}$, redshift in the range $0.2-0.4$ and declination $>-30^{\circ}$ \citep{Kale_2013A&A,2015A&A...579A..92K}. The occurrence of radio halos was found in $22\%$, of relics in $5\%$, and of mini-halos in $16\%$ of the EGRHS sample. A total of 31 upper limits for radio halos and 5 for mini-halos were reported using the method of injection of model radio halos in the cases of non-detections \citep{Kale_2013A&A,2015A&A...579A..92K}. Further to this survey, with the availability of SZ-detected clusters, nearly complete mass-limited samples have been surveyed. A scaling relation between the radio halo power and the integrated Sunyaev-Zel'dovich (SZ) effect measurements (host cluster mass) and the weakness of the bimodality in this plane was reported by \citet{2012MNRAS.421L.112B}. Targeted studies of Planck-discovered clusters also revealed new radio halos \citep[e. g.][]{2011ApJ...736L...8B,2015MNRAS.454.3391B}. Further, \citet{Cuciti_2021A&A} have surveyed 75 galaxy clusters from the Planck catalogue in the redshift range of 0.08-0.33 with masses and $M_{500} > 6\times10^{14}\rm{M}_{\odot}$. In this study, it was found that $\sim90\%$ of the radio halos are hosted in merging clusters, and their radio power is correlated with the mass of the host clusters, albeit with a large dispersion. Surveys of samples selected from the Massive Cluster Survey (MACS) and the South Pole Telescope have also revealed new cluster radio sources \citep{Parekh_2017MNRAS,Paul_2019MNRAS,2021MNRAS.500.2236R}. 
 
\begin{figure*}
    \centering
    \includegraphics[trim = 2.0cm 5.5cm 2.5cm 3cm, clip, width =14cm]{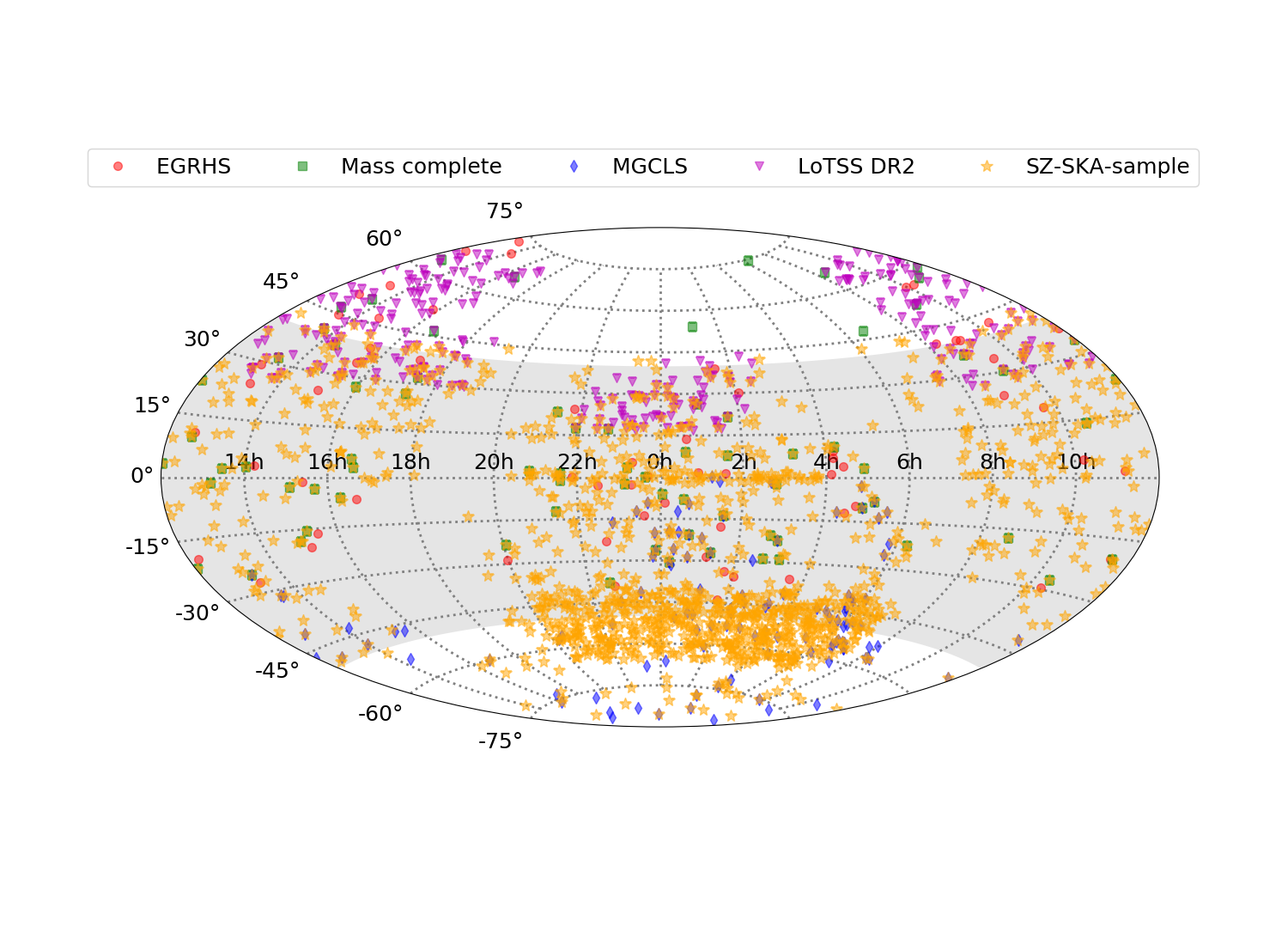}
    \caption{Various samples of galaxy clusters surveyed in the last decade are shown in an Aitoff projection of the sky. The red circles show the clusters in the Extended GMRT Radio Halo Survey \citep{2015A&A...579A..92K}, the green squares show the mass complete extension of EGRHS \citep{Cuciti_2021A&A}, the blue diamonds show the MeerKAT Galaxy Cluster Legacy Survey \citep{Knowles_2022A&A} and magenta triangles show the Planck clusters surveyed in LoTSS~DR2 \citep{Botteon_2022A&A}. The SKA will enable radio surveys of a large sample of galaxy clusters in the southern sky, such as the SZ-detected clusters from \citet{2017yCat..35940027P} shown by stars. The sky that is accessible to both the uGMRT and the SKA is shown in grey.}
    \label{fig:surveys}
\end{figure*}

MeerKAT, the precursor for the SKA in South Africa, has recently released the radio images of about 115 clusters at 1280 MHz, called the MeerKAT Galaxy Cluster Legacy Survey \citep[MGCLS;][]{Knowles_2022A&A}. This sample spans a declination range of $-85$ to $0$ degrees, a redshift range of $0.011<z<0.87$ and consists of X-ray and radio selected clusters based on published works and the MCXC catalogue. The first glimpse at samples of clusters at high redshifts ($>0.5$) was recently possible with the LOFAR. Among 20 clusters that were sampled, radio halos were found in $\sim50\%$ of the clusters, which imply that a process of fast magnetic amplification may be playing a role \citep{2021NatAs...5..268D}. The largest survey so far has been done recently by \citet{Botteon_2022A&A} taking a list of 309 clusters from the second catalogue of Planck SZ-detected sources that were in the area of 5634 deg$^2$ covered by the Second Data Release of the LOFAR Two-meter Sky Survey (LoTSS-DR2). The survey has median redshift of $z=0.280$ in the range $0.016 < z < 0.9$ and mass range of $1.1\times10^{14} M_{\odot} < M_{500} < 11.7 \times 10^{14} M_{\odot}$. With a median rms noise of $83~\mu$Jy, out of 309 clusters, 83 clusters are found to host radio halos, including candidate halos and 26 are detected with radio relics and candidate relics, with an overall detection rate of $30\pm11\%$ halos and $10\pm6\%$ relics, of which almost 50\% are new detection. This survey also includes the earlier survey of HETDEX spring region using LOFAR telescope \citep{Weeren_2021A&A}. We show the large surveys of clusters in the sky in Fig.~\ref{fig:surveys}. 
  
Targeted radio surveys with interferometers such as the GMRT require long observations and thus if suitable surveys of wide regions of the sky are available, the search for diffuse emission in galaxy clusters can speed up considerably. In Fig.~\ref{fig:surveys} we have marked the cluster sample from the Planck which will be accessible for studies with the SKA. There is a large overlap in the cluster samples accessible to uGMRT and the SKA between the declinations of $-50^{\circ}$ and $+40^{\circ}$ \footnote{The limiting declination is taken to be the one where the observing time is 3.5 h given the elevation limit of the respective telescope.}.

\section{Diffuse radio sources in clusters}\label{sec:cluster_radio_sources}

\subsection{Current classification of cluster radio sources}\label{sec:current_class}
\begin{figure}
\centering
\includegraphics[width=0.47\textwidth]{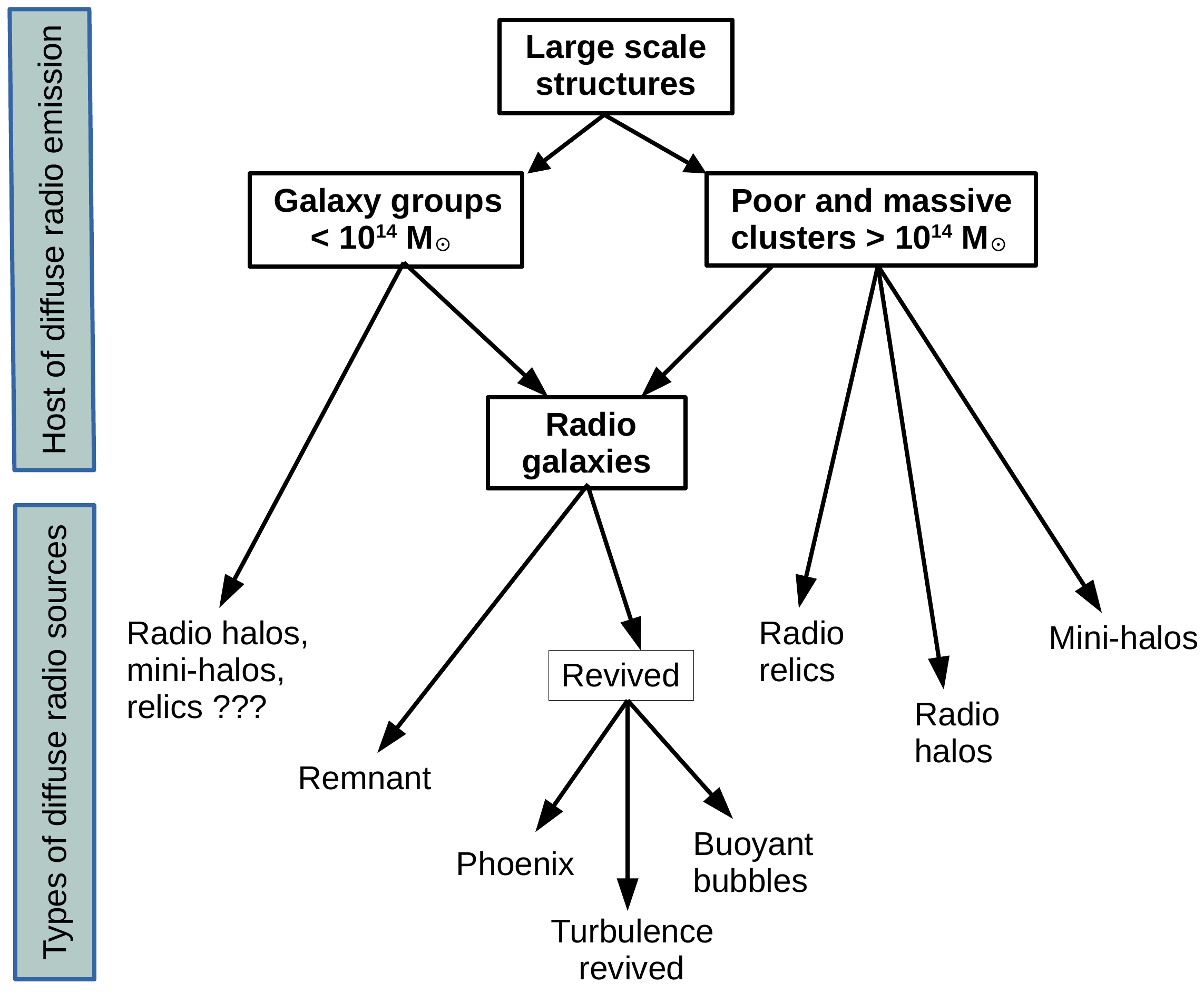}
\caption{Schematic diagram for the classification of cluster diffuse radio sources. 
}\label{fig:classification}
\end{figure}

With the growing statistics of the diffuse cluster radio sources, to ease systematic studies, one needs classifying these sources. Historically, these sources were classified broadly into Radio halos, Relics and mini-halos, based on a mixture of their location and the origin \citep{Ensslin_1998A&A,Giovannini_1999NewA,Feretti_2008LNP,Feretti_2012A&ARv}. Recently, with ample of data availability and with a better understanding of these sources, \citet{Weeren_2019SSRv} have made a new classification where they clubbed radio halo and mini-halos under the name (i) radio-halo, depending on their location and shape. Instead of radio relics, they named them (ii) cluster radio shocks, depending on their mechanism of origin, and finally, defined a new class of objects as (iii) revived AGN fossil plasma sources, phoenices, and GReET, because of their ultra-steep spectrum nature. Depending on the current classification schemes, we provide here a schematic diagram in Fig.~\ref{fig:classification}, and present the updates on these various diffuse radio sources in further subsections.

Although, radio halos, mini-halos and cluster radio shocks in general show a power-law spectrum, AGN relics or fossil plasma sources and phoenices show a rather curved spectrum due to the ageing effect. While the power-law spectrum favours an in-situ particle acceleration mechanism, spectral curvature indicates the dying or revival of aged synchrotron emitting electrons \citep{Weeren_2019SSRv}. It shows, that the distinct nature of these sources depends primarily on the initial spectrum of the electrons and the dynamics of the ICM. With the large number of discoveries of diffuse radio sources in and around galaxy clusters, a reconsideration of the classification may be needed.

\subsection{Radio halos and mini-halos}
Diffuse, mega-parsec sized radio sources that extend over the cluster volume, are more or less co-spatial with the thermal X-ray emitting gas of the ICM and cannot be associated with any of the cluster galaxies, are termed as radio halos. Radio halos typically have radio powers at 1.4 GHz in the range $10^{23} - 10^{26}$ W Hz$^{-1}$ and are found to be unpolarized \citep{Giovannini_2009A&A,Feretti_2008LNP}. The most powerful radio halos have been found in merging clusters \citep[e. g.][]{2001ApJ...553L..15B,Cassano_2010ApJ}. 
The radio halos are proposed to be powered by a re-acceleration process involving turbulence \citep[][for review]{2014IJMPD..2330007B} and the underlying population of relativistic electrons that are produced as a by-product of the hadronic collisions \citep{Dennison_1980ApJ, Schlickeiser_1987A&A}. Part of the mildly relativistic seed electron population may also be a result of the radio galaxies in the cluster. The theoretical models in this regard are described in Sec.~\ref{sec:theoretical_models_simulations}.

Radio mini-halos are diffuse radio sources surrounding the brightest cluster galaxies (BCGs) in relaxed, cool-core clusters, and have extents $\sim$100--500 kpc. Otherwise, they have observational radio properties almost similar to that of radio halos. Mini-halos are known to occur in both low and high redshift clusters, such as redshift 0.018 \citep[Perseus cluster,][]{sij93} and redshift 0.596 \citep[Phoenixcluster,][]{2014ApJ...786L..17V}\citep{raja2020apj}. A mini-halo has been recently reported in the cluster  ACT-CL J0022.2–0036 at redshift 0.8050 \citep{Knowles_2019MNRAS} and an intermediate radio halo in SPT-CL J2031-4037 \citep{raja2020}. 
The spectral indices ($\alpha$) 
of the few mini-halos that are known, are in the range of 1.2--1.6 at frequencies between 300--1400 MHz \citep{gia14,2020MNRAS.499.2934R}. The correlation between the BCG and MH radio luminosities  
indicates that the radio emission from the BCG itself (known to be AGN feedback due to multiphase cooling of cluster cores) may have a crucial role to play \citep{gia14}.
In a recent work, \citet{2020MNRAS.499.2934R} have studied a sample of 33 clusters with mini-halos and explored the connection between the Brightest Cluster Galaxy and the mini-halo. They suggest a connection between the feedback processes at the AGN and the mini-halo, in line with the earlier hypothesis by \citet{2016MNRAS.455L..41B}. 

The X-ray properties of clusters hosting mini-halos and radio halos are found to be very different. Mini-halos are found in relaxed clusters and radio halos in merging clusters \citep[e. g.][]{Cassano_2010ApJ,2015A&A...579A..92K}. The distribution of central entropy ($K_0$) in galaxy clusters shows that the clusters hosting mini-halos have $K_0<20$ keV~cm$^2$ and those hosting radio halos typically have $K_0>50$ keV cm$^2$ \citep{2017ApJ...841...71G}. 

The division between mini-halos and radio halos is getting blurred due to deeper observations leading to new discoveries that challenge the current classification.
Theoretical models have also proposed that the clusters that are intermediate to 
merging or relaxed may show the presence of diffuse sources with properties consistent with mini-halos as well as radio halos \citep{2014IJMPD..2330007B}. There are examples of transition systems such as RXCJ0232.2-4420 \citep{2019MNRAS.486L..80K} and also mini-halos that have a more extended component \citep{2014MNRAS.437.2163S,2017A&A...603A.125V} which is also steeper in spectrum \citep[e. g.][]{2018MNRAS.478.2234S,Savini_2019A&A,majid2021}. Classification as a mini-halo or a radio halo requires availability of X-ray data. In the recent search for diffuse radio emission in the LoFAR DR2 \citep{Shimwell_2022A&A}, the term mini-halo has been dropped to avoid the uncertain classification until X-ray data become available, and the term radio halo is used to describe centrally located diffuse radio emission in clusters \citep{Botteon_2022A&A}. A new look at the classification may be needed, in view of the variety of sources that are discovered.

\subsection{Radio relics (cluster radio shocks)}

Radio relics or cluster radio shocks are diffuse radio sources with elongated or arc-like morphology that typically occur at the peripheries of clusters, usually coinciding with the X-ray shocks. In some cases a pair of such sources occur on nearly opposite peripheries of clusters 
and these are termed as double radio relics \citep{Bonafede_2017MNRAS}. These radio shocks are found with sizes up to several Mpc, comparatively flatter spectrum than radio halos/mini-halos and are highly polarized having fractional polarization $p\sim10\textrm{--}50\%$ at 1.4 GHz; \citep{Weeren_2009A&A, Kierdorf_2017,A21082022, Rajpurohit+2022AnA, deGasperin+2022}.

Relics were proposed as originating in accretion shocks at cluster 
outskirts \citep[e. g.][]{1998A&A...332..395E}. However, evidence from morphologies and spectral indices suggests that the shocks are most likely merger shocks that travel outward with respect to the cluster centre \citep{Paul_2011ApJ,Nuza_2017MNRAS}. The Mach numbers of the underlying shocks are in the range 1--4 which are much lower than expected for accretion shocks \citep{2007PhR...443....1M}. Observational evidence has been found in a few clusters showing an overall consistency of radio and X-ray derived Mach numbers \citep{2013PASJ...65...16A, 2021A&A...654A..41R, Chatterjee_2022AJ} supporting the merger shock scenario. Moreover, the 
polarization studies also exhibit aligned magnetic fields, along with the elongation of the relics \citep[e. g.][]{2010Sci...330..347V,2012MNRAS.426.1204K,2021A&A...654A..41R}.

\subsection{Remnant radio galaxies}\label{sec:radio_remnant}

After the jets of radio galaxies cease to be active, the radio lobes evolve passively unless disturbed by any other 
event in the surrounding medium. Due 
to synchrotron radiation and inverse-Compton scattering, the relativistic electrons lose energy  
rapidly, making the radio lobe fainter. Over time scales of a million years or so, 
the lobe can be undetectable. Such diffuse radio sources can be found in clusters. 
The pressure of the intra-cluster medium can lead to confinement of the lobes and prevent expansion losses \citep{Ensslin_2001A&A}. 
Steep spectrum diffuse radio sources 
which have morphologies like double lobes but no obvious jet and a core can be classified as remnant radio 
galaxies \citep[e.g.][]{murgia_2011A&A, salunkhe_2022A&A}. A sample of such sources having steep spectra extracted from the NVSS and VLSS surveys was presented by 
\citet{2009ApJ...698L.163D}. Also, more recent studies with the Murchison Wide-field Array as well as MeerKAT (MGCLS) have resulted in discoveries of remnant 
radio galaxies \citep{2021PASA...38....8Q, 2021ApJ...909..198H, Oozeer_2021Galax,Knowles_2022A&A}.
However, such sources need not necessarily have steep spectra, as can be 
seen based on the properties of the remnant discovered with the LOFAR \citep{2016A&A...585A..29B}.  
Deeper surveys are likely to expand the samples of such remnants as we reach fainter levels of detection 
across the frequency range, such as covered by the SKA precursors and pathfinders. Systematic search in 
these images can result in a census of such sources that will throw light on the late stages of radio lobe evolution.

\subsection{Revived cluster diffuse radio sources }\label{sec:revived_sources}  

Numerous examples of another class of sources have been discovered, with the availability of highly sensitive low-frequency radio telescopes such as LoFAR and uGMRT, in recent times. They are of ultra-steep spectrum ($\alpha<-1.5$), mostly with high spectral age, with curved spectrum at high frequencies and are found anywhere inside the clusters. Usually, they are known as phoenices or AGN relics \citep{Weeren_2019SSRv}.
As the classification suggests, this diffuse emission come from the revival or re-energization of fossil non-thermal electrons in clusters, due to the passage of a shock or turbulence. The hosts of these sources are often the jets present in the cluster associated with active galactic nuclei (AGN), in their radio inactive phase. The spectral age of these jets or lobes are tens of Myr \citep{Ensslin_Gopalkrishna_2001A&A}. The faded synchrotron emission (see sec.~\ref{sec:radio_remnant}) from these sources are boosted by shock compression or gentle re-energisation or complex plasma interaction and appear bright in sub GHz radio band \citep{Ensslin_2001A&A,deGasperin_2017ScienceAdvances}. These sources typically have ultra steep integrated spectrum with $\alpha< -1.5$ \citep{Mandal_2019A&A}, shows steepening and curvature at high frequencies \citep{Slee_2001AJ} and irregular spectral index distribution \citep{Mandal_2020A&A}. They do not have specific location or shape based classification. The ultra steep nature makes them difficult to detect at GHz frequency, and thus these sources remain unexplored in detail. Though the low spatial resolution of the low-frequency sky surveys like TGSS, GLEAM, VLSS etc. makes it challenging to distinguish these sources from other diffuse structures, with the improvement of telescope facilities in recent times, several fossil sources have been discovered \citep{deGasperin_2017ScienceAdvances,Mandal_2019A&A,Mandal_2020A&A,Duchesne_2021PASA,Duchesne_2022MNRAS,Brienza_2022A&A}. Recent discoveries have helped in quantifying properties of these revived plasma sources and have significantly improved our understanding.
\begin{figure}
\includegraphics[width=0.47\textwidth]{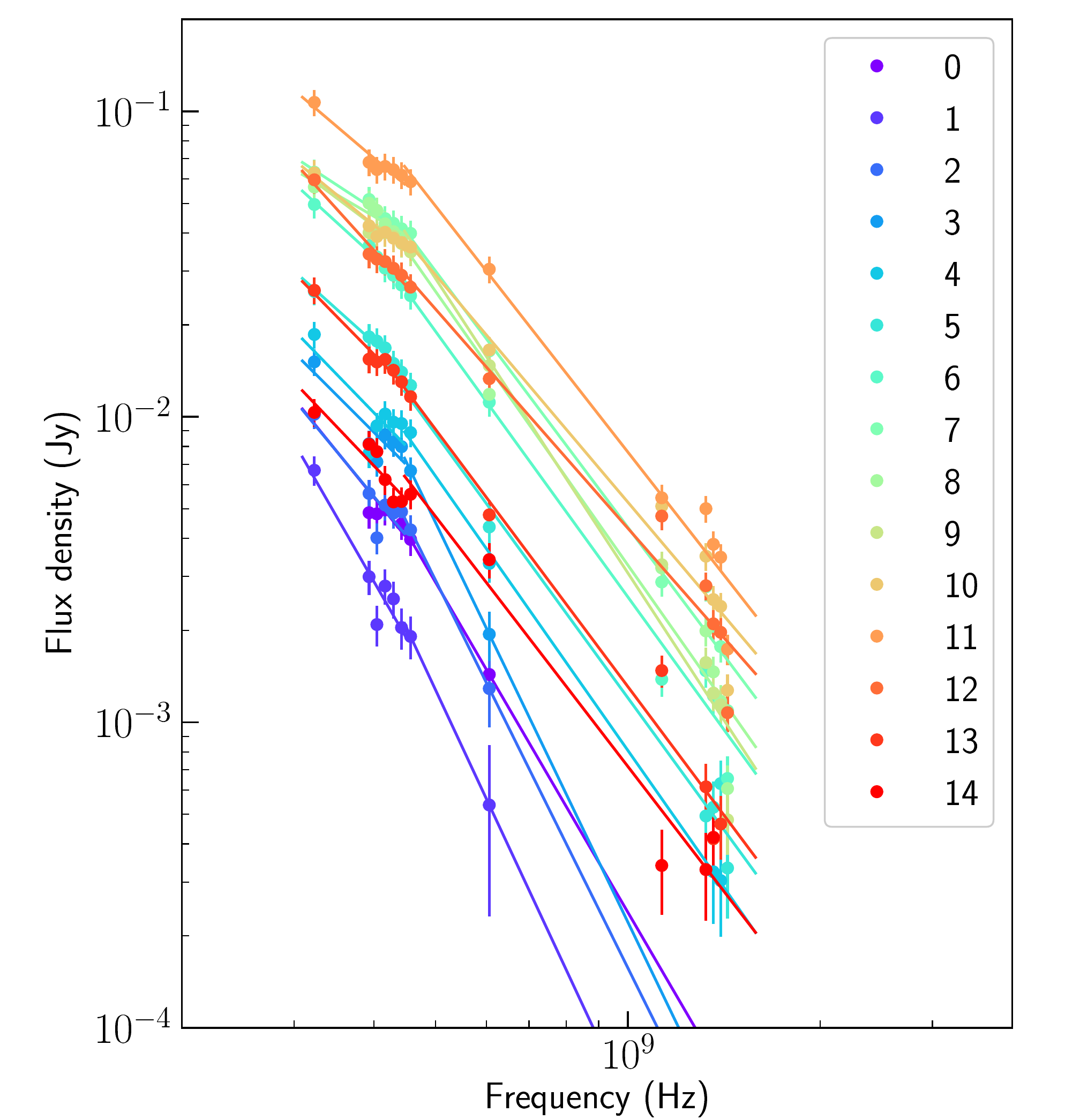}
\caption{The curved spectra of A4038 phoenix for different regions of the phoenix marked in different colours as mentioned in \citep{Kale_2018MNRAS}.}
\label{fig: phoenix_curvature}
\end{figure}

\subsubsection{Radio Phoenix:}
Although most of the diffuse radio sources  from the  peripheral regions of clusters 
have been traditionally classified as radio relics, some of these sources clearly have different physical origins, which led, \citep[e.g.][]{Kempner_2004rcfg} to suggest a different scheming of naming them based on our understanding of their origin. 
The radio phoenix is a class of diffuse source 
\citep[e.g.][]{Ensslin_2001A&A} where a plasma bubble resulting from past 
AGN activity is compressed by a passing shock wave (possibly from a recent merger). This provides sufficient energy 
to re-energize the relativistic electrons and also enhances the magnetic field strength.
This results into a low power steep-spectrum radio source, now classified as radio 
phoenix. Simulations \citep[e.g.][]{Enblin_2002MNRAS} have been used to find  likely candidates in radio observations. These sources can be both roundish (e.g., Abell~1664, \citealt{Kale_2012ApJ},  or elongated (e.g. Abell~2048, \citealt{Weeren_2009A&A}; A3017, \citealt{A3017-2021}), both in rich clusters (e.g. Abell~85, \citealt{Knowles_2022A&A, Rahaman_2022MNRAS.515.2245R})
or poorer systems (e.g. IC1262, \citealt{IC1262-2019}), and can be found in a range of cluster-centric radius, but closer to the core than in the case of the radio relics. Furthermore, unlike relics, 
radio phoenices are found in both merging and relaxed clusters.

\subsubsection{Turbulence revived ultra-steep spectrum sources:}

Turbulence, as well as a passing shock wave, can revive old fossil plasma in the ICM. The first case of this kind was observed in the cluster of galaxies Abell~1033, \citep{deGasperin_2017ScienceAdvances} 
where the tail of a WAT source gradually fades and then suddenly starts to brighten again. 
A spectral study of this source also shows the flattening of the spectrum in this part of the tail. 
A possible mechanism behind this source was proposed to be second-order Fermi acceleration by 
the turbulence generated by Rayleigh-Taylor and Kelvin-Helmholtz instabilities in the tail. 
Two other similar sources are found in ZwCl~0634.1+4750 \citep{Cuciti_2018A&A} and 
Abell~1314 \citep{Wilber_2019A&A}. Because of very few known examples of these sources, 
very little is understood about them. Future large scale surveys may provide a decent sample to work with.

\subsubsection{Buoyant radio bubbles:}
The X-ray image of a handful of cool core clusters has shown cavities near the central Brightest 
Cluster Galaxy (BCG), often coincident with buoyantly rising radio bubbles from the central 
AGN \citep{McNamara_2000ApJL,Fabian_2000MNRAS,Clarke_2005ApJ,Su_2017ApJ}. These buoyant radio bubbles interact 
with the ICM, inducing subsonic turbulence and offsetting the overall cooling of the gas. 
So far, no evidence of strong shock has been found near these bubbles \citep{Churazov_2013MNRAS}. 
Simulations have shown that the rise of the bubbles in the ICM undergoes Rayleigh-Taylor (RT), 
and Kelvin-Helmholtz instabilities, and their expansion is closely comparable to the 
mushroom clouds formed by massive explosions on earth \citep{Saxton_2001ApJ,Reynolds_2005MNRAS,Gardini_2007A&A}. 
The instabilities are balanced by viscosity and magnetic field \citep{Reynolds_2005MNRAS,Dong_2009ApJ,Ogiya_2018ARXIV}. 
The bubbles are less dense compared to the ambient medium, which gives rise to the buoyancy \citep{Blanton_2004RCFG}.
The gas displaced by the bubble is compressed in X-ray bright rings and remains in pressure 
equilibrium surrounding the bubbles \citep{Churazov_2002MNRAS}. Depending on their radio luminosity, 
they are classified as radio bright or radio-filled cavities (A2390, \citealt{Savini_2019A&A}), 
radio-dim or ghost cavities (Perseus Cluster \citealt{Fabian_2002MNRAS}); and intermediate cavities, 
where the radio emission partly fills the cavity (A4059, \citealt{Heinz_2002ApJL}). The short radiative 
lifetime of the cosmic ray electrons (only tens of Myr) makes it challenging to explore these sources in GHz frequency. Low-frequency radio observations with uGMRT, LOFAR can probe the radio emission from 
the old particle population of the AGN outflows in ghost and intermediate cavities.  
\citep{Russell_2019MNRAS,Birzan_2020MNRAS,Biava_2021A&A,Brienza_2021NatureAstronomy}. Deep radio observations of these sources are needed to establish radio x-ray correlations which would help in understanding  the bubble dynamics, the role of the magnetic field and also the feedback mechanism better. 

As we are approaching more and more towards lower frequency radio study with telescopes like  uGMRT, LOFAR, MWA, SKA-low (in future), many more details about these fossil sources are to be unveiled in coming years which would improve our understanding of these sources significantly.

\subsection{Magnetic fields in clusters from polarization studies}\label{sec:pol_icm}

Besides gravity, magnetic fields are also believed to play an important role in the evolution of the ICM.
A host of X-ray and radio continuum observations have revealed their physical properties,
such as, temperature, extent, presence of low Mach number shocks, and possible sites of particle acceleration 
\citep{1998A&A...332..395E, Feretti_2012A&ARv, Weeren_2019SSRv}.

A clear understanding of the structural nature and the coherence scale of magnetic fields in cluster halos is currently inadequate, and it is thus important to gather knowledge on the factors that drive turbulence in the ICM, and subsequently to  understand how they distribute and accelerate relativistic plasma, contribute to pressure balance, and affect the gas content in galaxies \citep[e.g.,][]{donnert2009, Vazza+2021_aap}. Much of our knowledge of the cluster magnetic fields comes from the diffuse synchrotron emission from cosmic ray electrons. This emission is expected to be partially linearly polarized \citep[e.g.,][]{govoni2013, SBS21}, 
and its measurement is capable of providing insights into the properties of the magnetic field structure in the ICM. 

Magnetic fields in the ICM have been mostly studied via analysis of the Faraday rotation measure (RM) of polarized radio
sources lying in the background or embedded within the magnetized ICM
\citep{Kronberg1994,CKB01, Murgia+04, Rudnick2004, Bonafede2010, BCK16}, the so-called \textit{RM-grid}. These studies show that the dispersion of RM ($\sigmarm$) decreases substantially as a function of impact distance from the cluster centre, where $\sigmarm \approx 200\radm$ towards the centre to $\sigmarm \lesssim 50\radm$ at distances $\gtrsim500\kpc$ from the centre. By comparing these results with numerical models, which assume Gaussian random distribution of magnetic fields in clusters, evidence for $\mkG$-level
fields that are ordered on scales of several $\kpc$ in the ICM have been found.

RM-grid-type studies are faced with a few challenges. Firstly, it is
difficult to disentangle the RM contributed by the ICM from that of the RM
intrinsic to the background sources. This might affect the inferred $\sigmarm$, and thereby an accurate understanding, of the magnetic field structures in the ICM \citep{Locatelli+2018}. Secondly, RM-grids are severely limited to only a few, $\mathcal{O}(10)$, background polarized sources per cluster \citep{Bonafede2010}, which makes constructing a 2-D RM-map arduous. Due to the limited number of Faraday RM measured towards background polarized sources of a cluster halo, the magnetic field along the line of sight in the ICM is discretely sampled and, therefore, background RM data from several clusters are statistically investigated \citep[e.g.,][]{BCK16}. Thirdly, while the assumption of a Gaussian random magnetic field distribution allows for
the power spectrum of the field to be expressed as a simple power-law and
subsequently apply to the observed RM data \citep[][]{CKB01, Murgia+04, Bonafede2010}, this assumption is in complete contrast to the
intermittent nature and non-Gaussian distribution exhibited by the field
components generated by fluctuation dynamos in the ICM \citep{Seta+20, SBS21}. Thus, a meaningful comparison between numerical predictions and observations of RM is
limited by assumptions made about the structure of magnetic fields in the cluster halo. This situation is unlikely to drastically improve, even with the advent of new and sensitive radio telescopes. For a proper quantification of the strength and structure of magnetic fields in the ICM, it is imperative to take into account the other two complementary observables that trace magnetic field structures ---
synchrotron emission and its polarization.

Despite years of work, polarized synchrotron emission from the radio
halos is yet to be convincingly detected. For a confident detection of
polarized emission from the diffuse ICM, the main problems are due to low surface brightness, strong
Faraday depolarization,
steep radio continuum spectra of the ICM, and confusion from polarized relic emission seen in projection. Based on cosmological magnetohydrodynamic (MHD) simulations, \citet{govoni2013} expect
that luminous halos may show a polarized flux density of 0.5--2 $\upmu$Jy at
1.4\,GHz when observed with 3\,arcsec angular resolution, corresponding to a
surface brightness of only 50--200\,$\rm nJy\,arcsec^{-2}$. Due to the large
Faraday dispersion expected in the ICM, \cite{BCK16} found that the presence of
a cluster medium increases $\sigmarm$ by about 60\,rad\,m$^{-2}$, and therefore, polarized halo emission
is expected to be severely depolarized at 1.4\,GHz. Going to higher frequencies
($\gtrsim2\ghz$) to reduce Faraday depolarization is also challenging, as the surface brightness of the synchrotron emission is further reduced due to the steep radio continuum spectrum ($\alpha \gtrsim 1$) of the ICM emission. 
Furthermore, since radio relics show substantial polarized emission near 1\,GHz \citep{Weeren_2010Sci, Wittor_2019MNRAS, Dominguez+2021}, even when observations are performed at relatively poor angular resolutions with single-dish telescopes
\citep{Kierdorf_2017}, the tentative polarized emission that have been reported for only three clusters, namely, for Abell\,2255 \citep{Govoni+05}, MACS\,0717.5+3745 \citep{Bona+09}, and Abell\,523 \citep{Girardi+16} are likely to be related to relics rather than the halo \citep{Pizzo+11, Raj+21}.
Hence, in order to identify an efficient observing strategy, it is necessary to investigate the properties of the polarized emission using various types of MHD simulations.

\subsection{High frequency radio observations}

The Sunyaev-Zeldovich effect produces distortions in the CMB spectrum through inverse Compton scattering of CMB photons off the energetic electrons present in and around cosmic structures, most notably galaxy clusters \citep{b99}. The SZe is proportional to the pressure (or energy density) of the electron population, and systematically shifts the CMB photons from the Rayleigh-Jeans to the Wien side of the spectrum. The SZe due to the thermal population of electrons such as in the ICM are well characterized through detections in directions of several galaxy clusters \citep{b99}. It is now well-known that the ICM does not comprise entirely of thermal plasma, but contains relativistic electrons and protons. Thus, the distortions to the CMB spectrum caused by all the populations of electrons (thermal and non-thermal) need to be accounted for. The generalized expression for the SZe including combinations of electron populations has been derived by \citet{c03}.

The Bullet Cluster is one of the most spectacular and well-known examples of an energetic cluster merger. It hosts a powerful radio halo in the central region \citep{malu2016}. The 5.5 GHz and 9 GHz detection of diffuse emission from the Bullet cluster has been performed with the ATCA. Due to limited short spacing, the images may suffer from missing flux from the diffuse emission. Similar high frequency work has been reported for Sausage Cluster at 16 GHz \citep{stroe2014}. 

Spectra of radio halos and relics are crucial for constraining models proposed to explain the emission. The spectra are expected to have one or more breaks, where the spectrum is expected to steepen. In addition to these breaks, basic parameters that characterize the energy distribution of the electrons in the atmospheres of the galaxy cluster are the high- and low–energy cut-offs in the spectra of diffuse emission, which consists of non-thermal populations of electrons. For characterizing the high-energy cut-offs, high-frequency observations of diffuse emission in clusters are clearly needed. Only two relics have been imaged in the frequency range 0.15– 30 GHz using a number of radio telescopes \citep{stroe2016}.

The energetics of non-thermal emission depends critically on the minimum energy to which the power-law distribution of relativistic particles is accelerated \citep{blundell2006}. This low-energy / low-frequency cut-off can be found in two ways, through observations of inverse Compton scattering (ICS) – through the ICS emission in X-ray, or through the SZ Effect \citep{basu16A&A, basu16Galaxy}. For non-thermal populations of electrons – as in radio relics – a detection of the SZ effect amounts to the most direct characterization of this low-energy cut-off. This has been confirmed for relativistic electrons in a radio galaxy jet, for the first time, by \citet{malu2017}, through their cm-wave observations. The SZ effect has also been recently used to detect warm-hot intergalactic matter (WHIM, \citet{deGraaff2019}) through stacking. The characterization of WHIM and low-energy cut-offs through the SZ effect makes a compelling case for high-frequency ($> 20$ GHz) observations of galaxy clusters – the two physical phenomena are promising techniques for probing the intergalactic/intracluster medium through the SZ effect. These are also relevant for SKA1-mid high bands up to 14 GHz. It is critical that SKA1-mid to retain sub-bands up to and above 14~GHz, which will aid to these high-frequency studies of galaxy clusters.

\section{Radio emission from poor clusters and groups}\label{sec:groups_n_low-mass} 
Structure formation in the Universe is expected to be self-similar on scales of supercluster filaments down to smaller systems of galaxies  
\citep[e.g.][]{Kaiser_1986MNRAS,Morandi_2016MNRAS}. However, the lower-mass end of these systems (i.e. poor clusters and groups)
have not been observed in as much detail as the richer systems, mostly because of the challenges in the observations (e.g. limitations in sensitivity) at various parts of the spectrum necessary for such studies using existing facilities. 
Earlier studies expressed the expectation that such self-similar properties would be true even for X-ray and non-thermal radio emission \citep{Vikhlinin_2006ApJ,Cassano_2013ApJ}. The emerging scenario from deeper observations and simulations indicates that such self-similarity might be broken in energy and mass scaling in lower-mass systems \citep[][and references therein]{Paul_2017MNRAS,John_2019MNRAS,Lovisari_2021Univ}.  

Low-mass clusters and groups are unstable to small mergers, owing to their shallow gravitational potential and are strongly affected by various non-gravitational processes 
\citep{Lovisari_2015A&A}. The content of cosmic rays in these systems is high, likely due to the influence of greater AGN feedback activity \citep{Gilmour_2007MNRAS,Li_2019MNRAS}.  Simulations on the evolution of turbulence and cosmic rays in low-mass objects show that these significantly deviate from the self-similar scaling found in richer clusters, having flatter slopes and higher fluctuations, indicating better visibility of a fraction of them through their non-thermal properties \citep{Paul_2017MNRAS,John_2019MNRAS}, such as the radio halo emissions \citep{Paul_2018arXiv}. This may also result in a systematic departure of these sources from the expected scaling relations. Furthermore, the low-mass clusters are usually cooler than the massive ones, the Mach number of a fraction of the merger shocks is expected to be noticeably higher inside the low-mass clusters \citep{Sarazin_2002ASSL}. This is because, unlike the thermal evolution, merger shock velocity does not directly depend on the mass of the host cluster. However, to examine these observationally, we should have enough data at the lower mass end, currently, which is largely missing in the available literature (see Fig.~\ref{fig:correlation_halos} and \S\ref{sec:low-mass}).

\subsection{Observations of low-mass clusters and groups} \label{sec:low-mass}

The environment of poor clusters and groups, dotted on the cosmic web, is the dominant location for the early evolution of galaxies, in contrast to the better-studied rich clusters, which contain only a few percent of galaxies \citep[e.g.][]{Freeland_2011ApJ}.  Galaxy groups are the dominant reservoir of baryons in the Universe, but their shallow gravitational potential is vulnerable to tidal disturbances, galaxy interactions and mergers.
Arguably, it is these low-mass systems that hold the key to the understanding of key processes of heating (feedback due to star formation or AGN activity) and cooling, the mechanisms and timescales of energy injection, and the effects of the galaxy and group evolution on the development of the hot intergalactic medium.  

For this, the basic observations are a combination of deep X-ray observations of the hot intergalactic medium and the radio observations of the synchrotron-emitting plasma associated with AGN activity and star formation. Coupled with these, observations of molecular and atomic gas and dust yield crucial information. 

 Deep X-ray observations from {\it Chandra} and {\it XMM-Newton} are being obtained for a wide range of system masses ($\sim10^{13}{\rm{M_\odot}}$ to a few times 
$10^{15}{\rm{M_\odot}}$) \citep{Popesso_2004A&A,Hofmann_2017A&A}, of which groups are generally referred to as systems with $\lesssim10^{14}{\rm{M_\odot}}$ \citep{Paul_2017MNRAS}, and poor clusters ($10^{14}{\rm{M_\odot}}\lesssim M \lesssim5\times10^{14}{\rm{M_\odot}}$).
For decades, VLA L-band and C-band observations revealed radio sources in central galaxies in low-mass systems \citep[e.g.][]{Burns1987}
The fact that diffuse radio emission was rarely detected in early shallow observations at L-band in such systems led to lack of information about activity of most nuclei in the central galaxies of low-mass systems. As AGN feedback began to emerge as the dominant source of heating in the intracluster gas in rich cluster, there were doubts about the nature of the  extra heating in the cores of clusters. However, deeper observations, particularly at lower frequencies initially with the GMRT, have revealed the presence of these sources in poor clusters such as AWM4 \citep{Osullivanawm42011}, and groups such as 
HCG~62 \citep{GittiH622010}, NGC~5044 \citep{DavidN50442009}, NGC~1407 \citep{GiacintucciN14072012}, NGC~741 \citep{NGC7412017} and many others \citep{Giacintucci2011,Osullivan2011}. In addition, as the relativistic plasma from these AGN become older, they lose energy and predominantly radiate at lower frequencies, thus being more prominent in low-frequency observations. Well-defined samples of galaxy groups need to be constructed for multi-wavelength observations in order to understand the systematics- the optically selected CLoGS study \citep{Clogs2022} or the X-ray selected XXL sample \citep{xxlradio2020} are positive steps in this direction.

The unprecedented sensitivity of LOFAR ($\sim80~\mu$Jy at 144~MHz) and uGMRT ($\sim20~\mu$Jy at band~3) has, however, revolutionized the study of diffuse emission in low-mass systems as observed through individual studies \citep{Hoang_2019A&A,Knowles_2019MNRAS,Botteon_2019A&A,Paul_2020A&A} as well as in a handful of systematic explorations \citep{Paul_2021MNRAS,Weeren_2021A&A,Botteon_2022A&A}. 

Although less numerous, low mass clusters are now detected with almost all types of diffuse radio sources, even the ultra-steep sources \citep{Mandal_2020A&A}. Particularly, the detection fraction of relics is more in low-mass systems than in high mass systems \citep[e.g.][]{Gasperin_2014MNRAS,Gasperin_2017A&A,Kale_2017MNRAS,Dwarakanath_2018MNRAS}. Almost 31\% of all diffuse emissions from low-mass systems are cluster-radio shocks (relics) compared to only 21\% in high mass systems, as observed by \citet{Botteon_2022A&A} in their large sample of 146 clusters with mass $M< 5\times10^{14}~M_\odot$ and 135 high mass systems from the Planck cluster list. So far, the lowest mass galaxy cluster known to host radio relics is Abell~168 \citep{Dwarakanath_2018MNRAS} with a mass $M\sim 1.24 \times 10^{14}~M_{\odot}$). From the correlations of radio halo power vs halo size and radio halo power vs mass \citep{Cassano_2007MNRAS, Feretti_2012A&ARv, Cassano_2013ApJ, Cuciti_2021A&Ab}, it can be inferred that radio halos detected in poor clusters are usually smaller in extent. However, a rare giant halo of size $\sim 750$ kpc has been detected by \cite{Botteon_2021ApJ} in one of the lowest mass ($\sim 1.9 \times 10^{14} M_{\odot}$) clusters PSZ2G145.92-12.53 using the LoFAR. It is also known to be the lowest power radio halo detected so far. Since low mass galaxy clusters are mostly dominated by the AGN activities, the dying component such as AGN relics with ultra-steep spectrum nature have been detected in some of them \citep[e.g.][]{Bruggen_2018MNRAS,Mandal_2020A&A}. \citep{Nikiel_2019A&A} even reported the detection of diffuse emission from the intergalactic medium in compact galaxy groups. With these discoveries, researchers have demonstrated that low-frequency telescopes like uGMRT and LoFAR have huge potential to detect diffuse cluster radio sources, even in low-mass to group scales.

\begin{figure}[h!]
\includegraphics[width=0.49\textwidth]{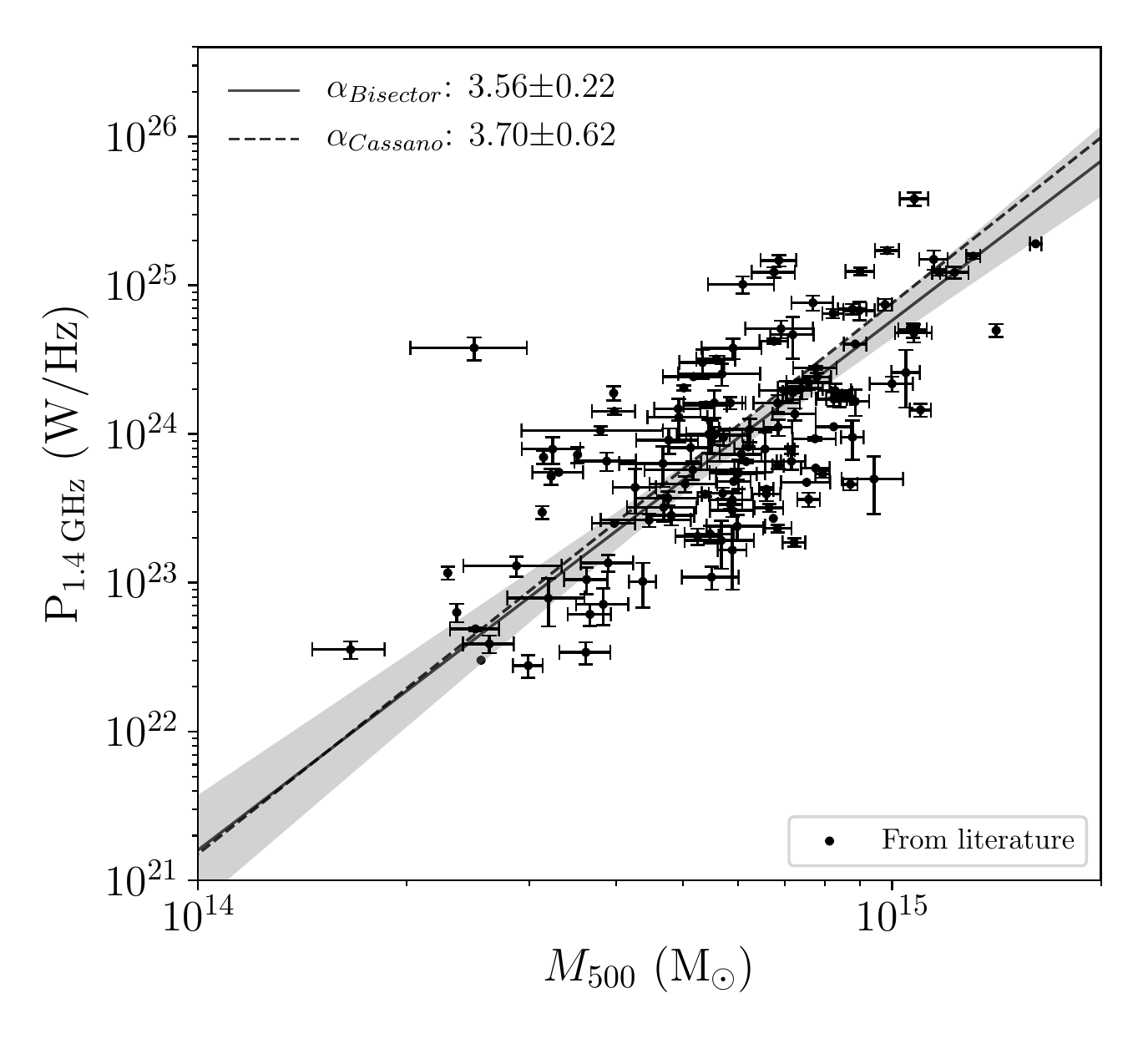}
\caption{Radio halo power plotted against mass $M_{500}$ of all the radio halos observed so far \citealt{Weeren_2019SSRv, Paul_2021MNRAS, Cuciti_2021A&Ab, Botteon_2022A&A, Hoang_2022A&A}. Radio power is scaled to 1.4~GHz, considering the spectral index as available in literature, otherwise the spectral index of $-1.3$ is assumed.}
\label{fig:correlation_halos}
\end{figure}

The existing radio halo power mass correlation ($P_{1.4~GHz}-M_{500}$) shows a steep spectrum (slope $\sim4$) and bi-modal distribution in Radio power--X-ray ($P_{1.4GHz}-L_{X}$) plane \citep{Cassano_2013ApJ}, predominantly with the massive objects. With the previous \citep[for a review][and references therein]{Weeren_2019SSRv} and more recent observations \citep{Paul_2021MNRAS, Cuciti_2021A&Ab, Botteon_2022A&A, Hoang_2022A&A} available in literature that revealed a sizable number of new radio halos and relics, some even in low-mass clusters \citep{Paul_2021MNRAS, Botteon_2022A&A, Hoang_2022A&A}, we re-plot and estimate the correlation for $P_{1.4GHz} - M_{500}$. Since, the  diffuse emission from many of the galaxy clusters have been observed in different radio frequency bands, to be consistent with the estimates reported in literature, the halo radio power is scaled to 1.4 GHz, assuming the spectral index of $-1.3$ (unless the spectral index is available in literature). The updated correlation plot is shown in Fig.~\ref{fig:correlation_halos}. The BCES bisector slope for $P_{1.4GHz} - M_{500}$ ($\alpha_{new}$ = 3.56 $\pm$ 0.22) by and large follow the \cite{Cassano_2013ApJ} correlation ($\alpha$ = 3.70 $\pm$ 0.56). However, the slight flattening of the slope, and a lack of data points below the mass $M < 4 \times 10^{14}$ M$_{\odot}$, motivates for more systematic and dedicated studies at the lower mass end. 

\section{Superclusters and the filamentary cosmic web}\label{sec:bridge-supercluster}

Superclusters of galaxies consist of galaxy clusters with linear or sheet-like inter-cluster filaments, corresponding to the denser crossroads of the cosmic web \citep[e.g.][]{Cautun_2014MNRAS} with groups of galaxies embedded in them. These are the most extensive and massive systems ($\geq$ 10$^{16}$ $M\odot$)  in the Universe that have decoupled from Universal expansion and are collapsing, significant parts of them being gravitationally bound. Cosmological n-body simulations and observations both support that strong concentrations of matter appear to be interconnected by vast, low-density filaments surrounded by volumes that are devoid of galaxies, which are identified as voids.

\subsection{Superclusters of galaxies}

In redshift surveys of galaxies and clusters, several well-known large-scale structures, that can be called superclusters according to the description above, have been identified. These include the Corona Borealis Supercluster  \citep{corbor1988}, the Coma Supercluster \citep{cfa1988,mahajan2011}, 
Hydra-Centaurus Supercluster \citep{hydcen1986},
Pisces-Cetus Supercluster \citep{pisces2005},
Hercules Supercluster \citep{hercules1979}, 
Shapley Supercluster \citep{shapley1989,Day1991,shapley1991}
Saraswati Supercluster \citep{2017ApJ...844...25B},
and our very own Laniakea Supercluster\citep{tully2014}, of which the Local group is a part, and many others that have been catalogued
\citep[e.g.][]{bahcall1984,2014MNRAS.438.3465T,Liivamagi2012}. Observations of superclusters are very important as they provide crucial information about the early evolution of structures in the Universe, and the formation and early evolution of galaxies, which spend most of their lives in this environment. 

One of the important aspects of studying such filamentary superclusters is to trace the large scale shocks that pass through the intra-cluster medium (ICM) (or intergalactic medium, IGM), and their effect on the formation of the complex cosmic web \citep{Miniati_2000ApJ,Ryu_2003ApJ,Pfrommer_2006MNRAS}.  Numerical simulations reveal different kinds of shocks, which control the overall evolution of large scale structure. These mainly belong to two categories: `internal' and `external' shocks. The former (internal or merger shocks) affect the material that is already heated to approximately the cluster temperature ($\sim$ 10$^{7}$ K) and may be responsible for generating cluster scale ($\sim$ 1 Mpc) diffuse radio sources in the form of halos and relics (mostly in massive clusters). The latter kind (external or accretion shocks)  is responsible for heating the cold gas surrounding the cluster out to the virial radius, and may be associated with infall of matter along the filamentary web surrounding the cluster \citep{Hoeft_2007MNRAS,Paul_2011ApJ}. 

The merger shocks have a low Mach number ($\mathcal{M}$ $\sim$ 2--3), while accretion shocks have a high Mach number ($\mathcal{M}$ $\sim$ 10--10$^{2}$). Such infall velocities are high enough that the infalling material can be accelerated to total energies of 10$^{18}$-10$^{19}$~eV while accreting onto collapsing structures. In the presence of even a weak magnetic field ($10-100$ nG) where the energy density of the magnetic field accounts for only $1\%$ of the total post-shock energy density, it is possible to detect the radio synchrotron emission that couples with accretion shock \citep{2004NewAR..48.1281W}. Thus, sensitive and high-resolution radio observations of superclusters can be useful to probe the underlying distribution of cosmic rays and magnetic fields \citep{vernstrom2017,Miniati_2000ApJ}. 

Simulations \citep{2015A&A...580A.119V, 2012MNRAS.423.2325A} predict that it is possible to detect extended diffuse radio sources associated with filaments if the shocks accelerate the electrons sufficiently \citep{2004ApJ...617..281K}. \citet{2018MNRAS.479..776V} have already detected, using the Sardinia Radio Telescope at L-band, a new population of faint and diffuse 28 candidate radio sources associated with a large scale filament.  They have identified nine massive clusters ($z$ $\sim$ 0.1) surrounding filaments in their survey area, which covers 8$^{\circ}$ $\times$ 8$^{\circ}$. The sizes of these radio sources vary from 0.3 to 8.6 Mpc. The mean radio power and mean X-ray luminosity of these new radio sources are $10-100$ times lower than those of diffuse radio emissions normally associated with clusters. Moreover, superclusters are promising targets to attempt the detection of the very challenging warm-hot intergalactic medium (WHIM) in radio bands which shed light on shock heated collapsing diffuse intergalactic medium (IGM) in filaments \citep{2000ApJ...534L...1T,1999ApJ...514....1C}. In order to accurately map the ``Cosmic web'' after subtracting the confused foreground or background compact radio sources, we need high surface brightness sensitivity ( $<\mu$Jy arcsec$^{-2}$) and sub-arcsec resolution of the SKA.  

\begin{figure}[h!]
\includegraphics[width=0.5\textwidth]{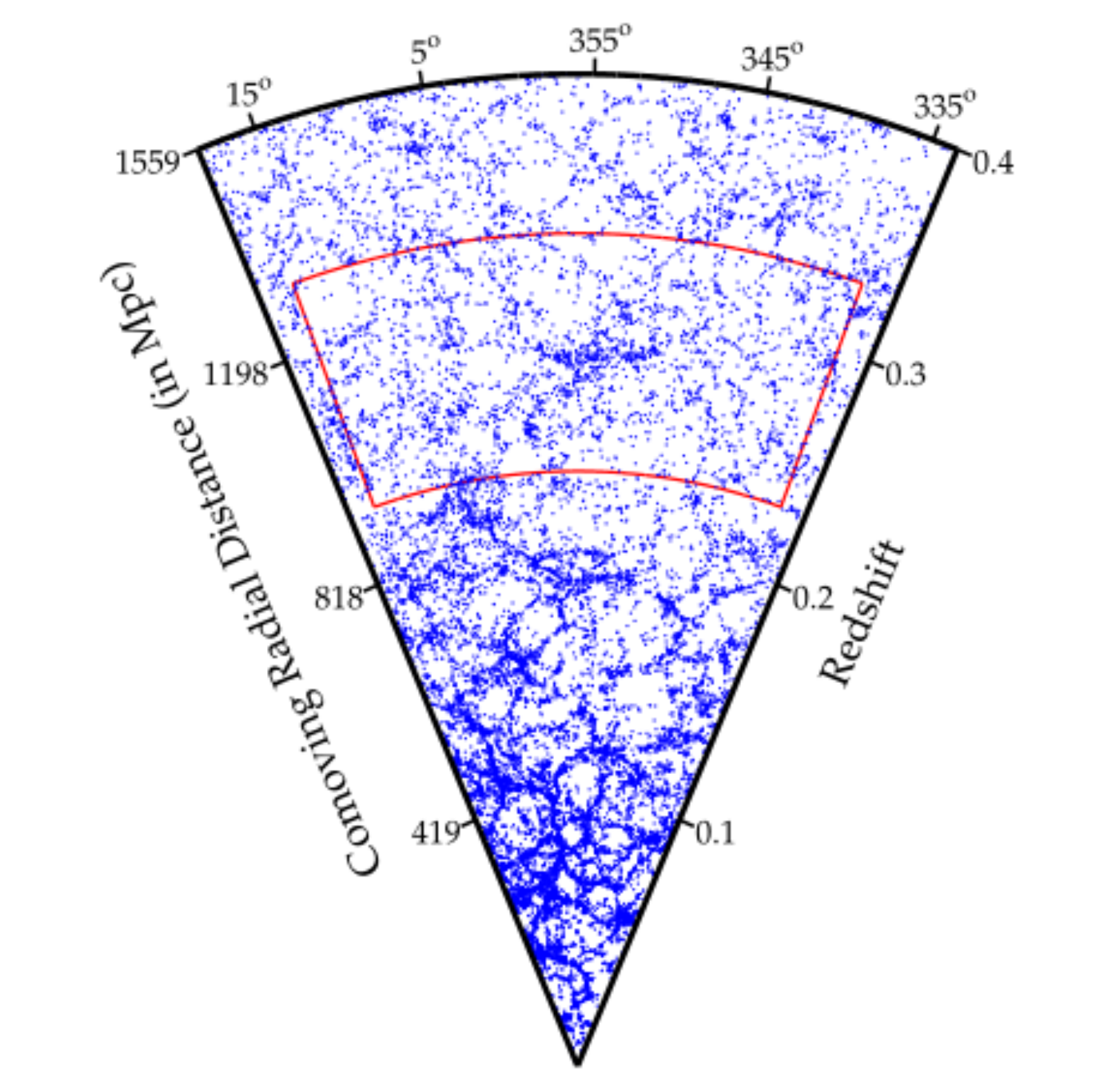}
\caption{Redshift cone plot for the {\it Saraswati} supercluster contains SDSS galaxies from $z$ = 0--0.4. The red box shows the concentration of galaxies of the {\it Saraswati} supercluster of width in RA 45$^\circ$ and Dec 2.5$^\circ$ (which is not shown for clarity). This figure is taken from \cite{2017ApJ...844...25B}.}
\label{sar_img}      \end{figure}

As a part of the radio observations of a supercluster with future generation radio telescopes, we are studying one of the extensive superclusters {\it Saraswati} with MeerKAT and uGMRT. The {\it Saraswati} supercluster was identified in the stripe~82 SDSS region, at redshift $z$ $\sim$ 0.3 \citep{2017ApJ...844...25B}. The total mass and size of this supercluster are $\sim 2 \times 10^{16}\,M\odot$ and $\sim 200$ Mpc, respectively, making it one of the largest observed structures in the Universe. Figure~\ref{sar_img} shows how SDSS galaxies are distributed in the {\it Saraswati} supercluster in a redshift cone plot (taken from \cite{2017ApJ...844...25B}). As seen in this plot, the {\it Saraswati} supercluster is situated (red box) within 336$^\circ$ $<$ Ra $<$ 16$^\circ$, -1.25$^\circ$ $<$ Dec $<$ 1.25$^\circ$ in the 270 degree$^2$ equatorial Stripe 82 region. In sensitive MeerKAT L-band pilot radio observations of the {\it Saraswati} supercluster \citep{2022MNRAS.509.3086P}, our primary aim is to detect faint and diffuse radio sources associated with filaments and study their properties. 

\subsection{Radio ridges in interacting clusters}
\par Apart from radio halos and relics, as described above, due to the filamentary nature of the cosmic web, large-scale diffuse radio sources are also detected beyond the main extent of galaxy clusters, at the outskirts where the electron density is very low as compared to the cluster core. These rare, steep spectra 
and low-surface brightness radio sources could be well detected with low-frequency radio observations. Such ``radio ridges''  have now been detected in only a few systems, as filaments between pairs of merging clusters \citep{botteon2018lofar, govoni2019radio,Botteon_2020MNRAS,Venturi_2022A&A}. 

The first low surface brightness radio ridge of this kind was found in the Coma supercluster, connecting the central 
halo and the outer relic \citep{kim1989discovery}. The corresponding X-ray observation (of the Coma cluster) suggests 
the moderate correlation between X-ray and radio brightness of the Coma bridge. 
Also, The effect of the flow of the cold diffuse gas down the filaments connecting the clusters in the supercluster can be seen in the star formation properties of galaxies in the Coma and A1367 clusters\citep[e.g.][]{Seth2020}. This further indicates that the X-ray
and radio emission is generated from the same volume and there is no projection effect. 

Among the most notable discoveries in this genre
is the spectacular $\sim$5~Mpc size radio ridge connecting the two massive clusters A399 and A401 \citep{govoni2019radio}, 
discovered with the Low-Frequency Array (LOFAR) \citep{van2013lofar} at 144 MHz frequency. 
Previously, X-ray observations of the A399-A401 galaxy cluster have revealed a hot (6-7 keV) X-ray filament between the pair 
\cite{2008PASJ...60S.343F}. Moreover, \citep{murgia2010double} detected radio halos with each of the clusters at their centre. 

The detailed morphology of the cosmic web shows rich large-scale structures of interconnecting filaments with clusters and groups embedded in them, and the interaction of warm and cold gas on these filaments is likely to have a profound influence on the regulation of star formation in the early life of galaxies \citep[e.g.][]{Pimbblet2004,Mahajan2012}.
In the core of the Shapley Supercluster, \citet{Breen1994} had pointed out a chain of four merging clusters (the central cluster A3558, group SC1327-312, group SC1329-313 and A3562, from Einstein and ROSAT observations.
 Recently \cite{Venturi_2022A&A}, using high sensitive MeerKAT L-band data, reported the first GHz detection of inter-cluster diffuse emission ($\sim$1~Mpc scale) joining A3562 cluster and group SC1329-313 along this filament. This bridge had not been detected in previous radio observations, only being detected in the deep MeerKAT data. This certainly points to further possibilities of direct discovery in supercluster filaments along the cosmic web filaments. 
Another spectacular example is that of  ZwCl 2341.1+0000, a complex merging structure of galaxies located at $z=0.27$, where a system of three merging clusters along with radio and X-ray emitting ridges are found, along with post-merger relics, as the heart of the Saraswati Supercluster
\citep{Zw2002,Weeren_2009A&A}.

\par The origin of radio synchrotron emission from these radio ridges is not well understood. The radiating timescale of electrons 
at 140~MHz is short, and they can travel only a distance $<0.1$~Mpc in their lifetime. On the other hand, the sizes of radio ridges 
are $\gtrapprox$ Mpc, which suggests {\it in situ} particle re-acceleration mechanism required for the radio ridge emission. Simulations 
have revealed a dynamically complex structure of radio ridge in which a shock acceleration mechanism, due to a cluster merger, is not the main source of the observed radio emission \citep{govoni2019radio}, as is the case with radio relics. It is rather a turbulence in the inter-cluster bridge, caused by complex substructures embedded between galaxy clusters. This turbulence produces weak shocks ($\mathcal{M} \le 2-3$) that cause re-acceleration of the pre-existing fossil electrons of energies ($\sim$ GeV), which interact with the ($\sim$nG) magnetic fields and generate radio emission. \citet{brunetti2020second} have predicted that the Fermi(II) re-acceleration mechanism is most favoured under these physical conditions and gives rise to long dynamical time scales of radiating electrons in the bridges connecting clusters. Observing this type of radio object is a major challenge due to its truly diffuse nature on such large spatial scales. We are interested to observe X-ray selected binary merging clusters (which also have hot X-ray filaments) with new radio telescopes such as MeerKAT and uGMRT to detect the radio ridges and probe their properties \citep{2017MNRAS.470.3742P,2019MNRAS.tmp.2668P,2020MNRAS.499..404P}.

\section{Theoretical models and efforts on simulating cluster radio emissions and magnetic fields}\label{sec:theoretical_models_simulations}

The diffuse radio emissions found in the cluster medium are theoretically understood to be generated due to synchrotron radiation from GeV electrons in a magnetized medium (\citet{Giovannini_2000NewA} and for review \citet{Ferrari_2008SSRv, 2014IJMPD..2330007B}, and references therein). Therefore, the detection of these diffuse radio structures in clusters provides indirect evidence for the non-thermal component of cosmic rays as well as the magnetic field in the ICM (e.g., \cite{Large_1959Natur,Willson_1970MNRAS}). Cosmic rays are usually the charged particles re/accelerated due to the merger driven shocks and the turbulent flows in the intracluster medium (ICM). The particle (re-)acceleration supposedly happens either from the ICM thermal pool, pre-existing electrons or the radio quite AGN lobes through two well-known processes, namely Fermi acceleration of orders one and two. Although the entire processes are difficult to unfold directly from observations, in recent decades, the cosmological simulations of clusters of galaxy formation are a leap forward in understanding the processes quantitatively. 

\subsection{Diffusive shock acceleration (Fermi-I)}
The diffuse radio emission at the periphery of the galaxy cluster has a morphology similar to the shock fronts in cluster simulations, \citep{Ryu_2003ApJ, Skillman_2008ApJ, Vazza_2009MNRAS, Paul_2011ApJ, Skillman_2011ApJ} which closely correlate the radio relics to the merger shocks. At the shock fronts, a small fraction of thermal electrons gets accelerated to relativistic energies, known as the diffusive shock acceleration mechanism (DSA), which in turn emits synchrotron radiation. In the DSA process, the charged particles temporarily get trapped inside the shock region and gain energy each time they reflect back to the upstream (pre-shocked) region across the shock \citep{Drury_1983RPPh, Blandford_1987PhR, Jones_1991SSRv, Malkov_2001RPPh}. The magnetic field perturbations induced by plasma effects in shock-medium, accelerate the electrons by reflecting back to the upstream, this process is also known as the first-order Fermi mechanism. These accelerated electrons in the presence of compressed magnetic fields emit synchrotron radio emissions along the shock surface and its downstream, which can successfully explain the elongated diffuse radio structures at clusters scales, known as the radio relics (as discussed above). According to \cite{Drury_1983RPPh} and \citep{Blandford_1987PhR}, for a planar steady shock, the electrons that are accelerated (via Fermi first-order process) by the shock of Mach number $\mathcal{M}$ form a power-law distribution in momentum space;
\begin{equation}
f(p) \varpropto p^{-q}	\;\;\;\;\;\;\;\;  ;  \;\;\;\;\;\;\;\; q = \frac{4\mathcal{M}^2} {\mathcal{M}^2 -1}
\end{equation}
With this, the radio synchrotron spectrum due to these accelerated electrons follow the power-law $S_{\nu} \varpropto \nu^{-\alpha_{inj}}$, with spectral index;
\begin{equation}
\alpha_{inj} = \frac{\mathcal{M}^{2}+3}{2(\mathcal{M}^{2}-1)}
\end{equation}
while the volume integrated radio spectrum becomes $S_{\nu} \varpropto \nu^{-\alpha_{int}}$, where $\alpha_{int} = \alpha_{inj} + 0.5$ \citep{Ensslin_1998A&A, Kang_2011JKAS}. The observed radio spectral index at the shock surface is often compared with the theoretical spectral index to infer the Mach number of the shocks.

The particle acceleration efficiency is constrained to the particle acceleration from the thermal pool. Recent studies by \cite{Botteon_2016aMNRAS, Eckert_2016MNRAS, Hoang_2017MNRAS} indicate that the particle acceleration from the thermal pool requires the large acceleration efficiencies to produce the total radio luminosity of the radio relic. Again, \cite{Botteon_2020A&A} in his study, tested the scenario with ten well-studied radio relics where shocks were observed in X-rays as well. They calculate the electron acceleration efficiency of these shocks if injected from the thermal pool, to reproduce their observed radio luminosity. And report that the standard DSA model cannot explain the origin of the relics if efficiency is smaller than $10\%$, as constrained by the studies of SN-driven shocks in galaxies. Whereas, SN-driven shocks are very strong with a high Mach number ($>10^{2}$) and are also found in a low beta-plasma ($\beta$ = gas pressure/ magnetic pressure). In contrast, the shocks in ICM are much weaker where ICM plasma has high beta $\sim 100$ \citep{Kang_2014ApJ}.

\subsection{Shock detection in cosmological simulations}
Cosmological simulations are essential to closely study shocks resulting from cluster mergers. Identifying shock structures in simulations is crucial but significant to understanding the kinetic energy flux mediation through shocks in the ICM. A fraction of the shock kinetic energy thermalizes the ICM, and helps in achieving the virialization state. The remaining energy results in the production of cosmic-ray particles (protons or electrons) through the first-order Fermi mechanism.

In the literature, there are a few grid-based methods \citep{Miniati_2000ApJ, Ryu_2003ApJ, Skillman_2008ApJ, Vazza_2009MNRAS, Vazza_2012MNRAS} and in SPH code \citep{Pfrommer_2006MNRAS, Hoeft_2008MNRAS} that identify the shocks and its magnitude, i.e, the associated Mach number, using the Rankine-Hugoniot jumps conditions across the shocks. \cite{Miniati_2000ApJ} $\&$ \cite{Ryu_2003ApJ} attempt to identify the shock in the `single-grid' codes and estimate the shock Mach number using temperature jump condition. They marked the shocks as, 'accretion shocks', where for the first time the photo-ionized gas is shock-heated and 'merger shocks', where the pre-shock gas has previous shock encounter(s). In addition, \cite{Ryu_2003ApJ} accounts for the floor temperature of intra-galactic medium (IGM), i.e. the minimum gas temperature, $T_{min} \approx 10^{4}$~K, created by the re-ionization of stars. This prevents us from over-estimation of the accretion shock strength in adiabatic simulations. The obliquity of the shock, multiple shock cells, and directional dependence in the shock cells are the significant issue to address. \cite{Ryu_2003ApJ} adopts the co-ordinate split approach to estimate the shock Mach number, whereas \cite{Skillman_2008ApJ} determine the shock strength in the direction of shock propagation. \cite{Skillman_2008ApJ} made the first attempt to study the shock in adaptive mesh refinement grid-based code where the former attempts are on uni-grid-simulations, and report that the previous methods \citep{Ryu_2003ApJ} overestimate the number of low Mach number shocks by a factor of $\sim 3$ due to the misconception about the direction of shock propagation. Later, \cite{Vazza_2009MNRAS} compute the shock strength using the velocity jump criterion which is very well consistent with the temperature jump criterion, except, the velocity jump method is more reliable at the outskirts of galaxy clusters or low-density environment in estimating the Mach number. All above-discussed shock detection schemes identify the shock and its strength as a post-process, which may cause uncertainties in the characterization of shocks since they consider the ideal conditions across the non-shocked-cells, i.e., no velocity and no temperature gradients, which already exist in a complex cosmological flow (in velocity and temperature fields) wherein the thermodynamic gradients due to shock are superimposed \citep{Vazza_2009MNRAS}. It demands the run-time shock detection to overcome the uncertainties, \cite{Vazza_2012MNRAS} attempt to hastily identify the shock, in a grid-based adaptive-mesh refinement (AMR) simulation code (ENZO), moreover, introduce a method to increase the level of refinement at the locations of shock fronts. Where, the AMR criterion based on one-dimensional velocity jump is added to the usual AMR criteria as adopted for gas and dark matter over-density in \cite{OShea_2004astro.ph}. Additionally, in \citep{Vazza_2012MNRAS} the total shock strength is estimated by the pressure jump conditions which account for the feedback from cosmic-ray flux at the shock fronts. Even hydrodynamical cosmological simulation basically describe the merger shocks, achieving sufficient numerical resolution, and locating the shock fronts and their strength is still a challenge.

\subsection{DSA in cosmological simulations}
\begin{figure}
	\includegraphics[width=\columnwidth]{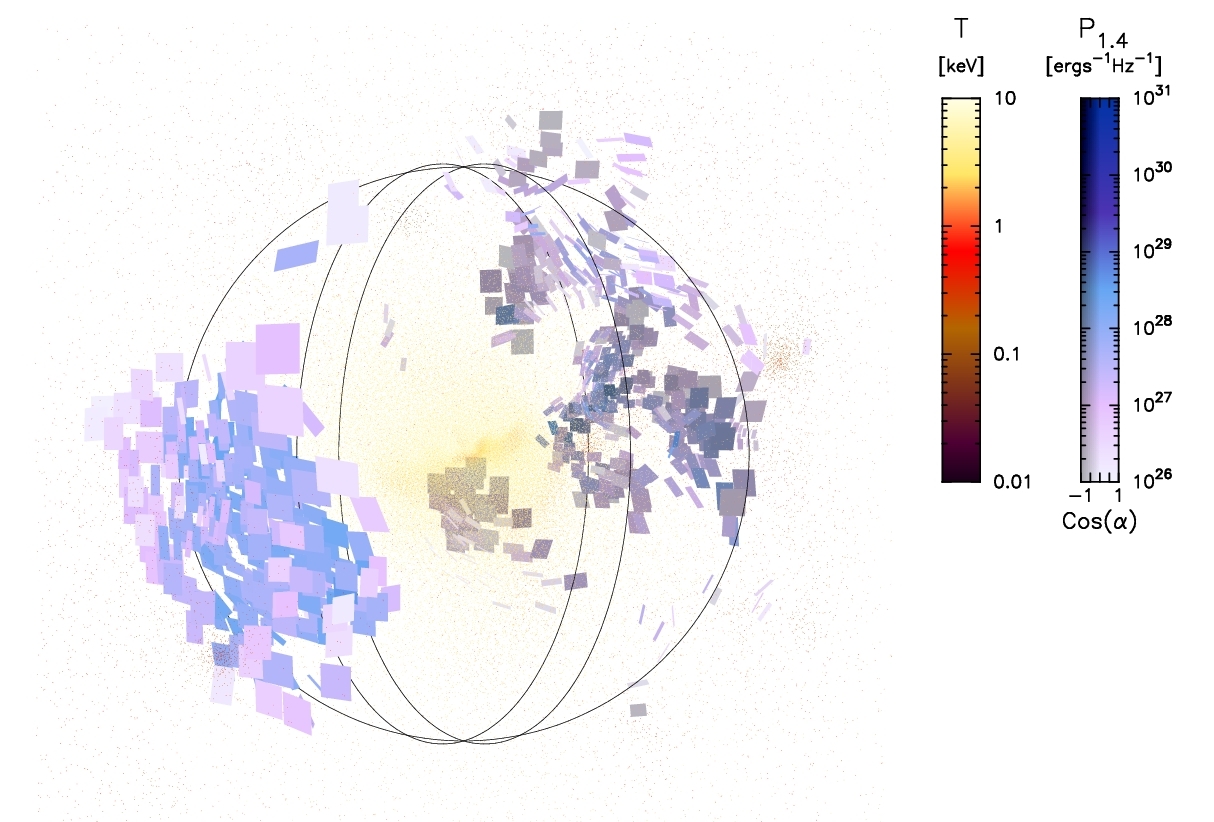}
    \caption{This is an example of double relic (Figure taken from \protect\cite{Hoeft_2008MNRAS}), representing two large shock fronts on the cluster periphery, from the SPH-MareNostrum Universe simulation. This shows the spherical nature of shocks, very similar to the radio relics found in Abell~3667 and Abell~3376.}
    \label{fig:DSA_inSPH1}
\end{figure}

In the last decade, many attempts were made to estimate the synchrotron radio emission from the shock waves \citep{Hoeft_2007MNRAS, Skillman_2011ApJ, Hoeft_2011JApA, Nuza_2012MNRAS, Nuza_2017MNRAS, Paul_2018arXiv, Wittor_2019MNRAS, Bruggen_2020MNRAS, Paul_2020A&A, Wittor_2021MNRAS, Wittor_2021arXiv}. The widely used semi-analytical solution to compute the total monochromatic radio power at frequency $\nu$ from a shock wave of area $A$, having electron efficiency $\xi_{e}$, with the downstream temperature $T_{d}$, the downstream electron density $n_{d}$, and magnetic field strength $B$, have been derived by \citep{Hoeft_2007MNRAS} as,
\begin{equation}
P(\nu) \varpropto A\; n_{d}\; \xi_{e}\; \nu^{-\alpha_{\mathsf{int}}}\; T^{\frac{3}{2}}_{d}\;\frac{B^{1+\alpha_{\mathsf{int}}}}{B^{2}_{\mathsf{CMB}} + {B}^{2}}
\label{eq:DSA}
\end{equation}
where $B_{\mathsf{CMB}}$ is the magnetic field corresponding to the energy density of CMB. Eq.~\ref{eq:DSA} enables one to implement it as a post-process to estimate the radio power in any cosmological/galaxy cluster simulation (e.g. Fig.~\ref{fig:DSA_inSPH1}). The underlying assumptions to derive this relation is that the electrons (i.e., with minimum momentum, say $p_{min}$) from the thermal pool, having Maxwellian distribution, get accelerated by the shock to a power-law distribution which is closely related to the Mach number of the shock (DSA theory). These accelerated electrons emit in radio through synchrotron process in a magnetized medium.

If one performs magneto-hydrodynamic simulations, the magnetic field strength as direction is the associative property of the cell in grid simulations (e.g. ENZO), else the approach will be to govern the magnetic field strength using semi-analytical methods \citep[e.g][]{Roettiger_1999ApJ}. The magnetic field strength is related to the baryonic matter density, i.e., modelling magnetic field strength with density \citep{Bonafede_2009RMxAC, Skillman_2011ApJ} and modelling the magnetic field density with the turbulence energy density of the medium \citep[e.g.][]{Subramanian_2006MNRAS, Beresnyak_2016ApJ,Paul_2018arXiv}. \cite{Paul_2018arXiv} demonstrate that the magnetic field in the ICM using magnetic energy density's sub-equipartition with turbulence energy density successfully explains the cluster scale magnetic field. They compare the magnetic field radial profile of the simulated coma-cluster like environment, which is in good agreement with the radial profile studied using the Faraday rotation measure (RM) \citep{Bonafede_2009RMxAC}.

Still, the DSA mechanism is poorly understood. The bottleneck is the minimum momentum required for the thermal electron to accelerate it to relativistic, non-thermal distribution and the efficiency of the shock to accelerate the sufficient charge particles for radio emission.

There are some more puzzles yet to address in DSA. If every shock can produce relativistic electrons from the thermal pool, then correspond to the shocks for every merging cluster there should be the associated radio relics. However, only a few per cent of merging clusters host the radio relics \citep{Feretti_2012A&ARv}. Furthermore, not all X-ray detected shocks show the signature of diffuse radio relics. Moreover, in some relic structures, the strength of the X-ray Mach number computed from the density/temperature jump does not agree with the radio-Mach number estimated from the radio spectral index. In order to address these puzzles, several authors \citep{Kang_2011ApJ, Kang_2012JKAS, Pinzke_2013MNRAS, Kang_2016ApJ} proposed the re-acceleration model in which the seed electrons are not from the thermal pool, instead they are considered to be pre-existed in the ICM as fossil electrons as the leftover of radio jets or lobes of the radio-quiet AGN. These radio-quiet electrons may be reborn as radio phoenix if re-accelerated by the merger shocks (also known as the adiabatic compression process) \citep{Ensslin_2001A&A}. The re-acceleration model is quite consistent with the observed radio relics where the X-ray Mach number shows a discrepancy with radio spectra-derived Mach number (e.g. Toothbrush radio relic,\citep{Kang_2016ApJ}. This confirms the strong connection between the fossil electrons from AGN and shock acceleration in ICM, provided the cluster must host an AGN (radio-quiet?) like in the case of Toothbrush radio relic \citep{Weeren_2016ApJ}.

\subsection{Turbulence Re-acceleration Mechanism}
\begin{figure}
	\includegraphics[width=\columnwidth]{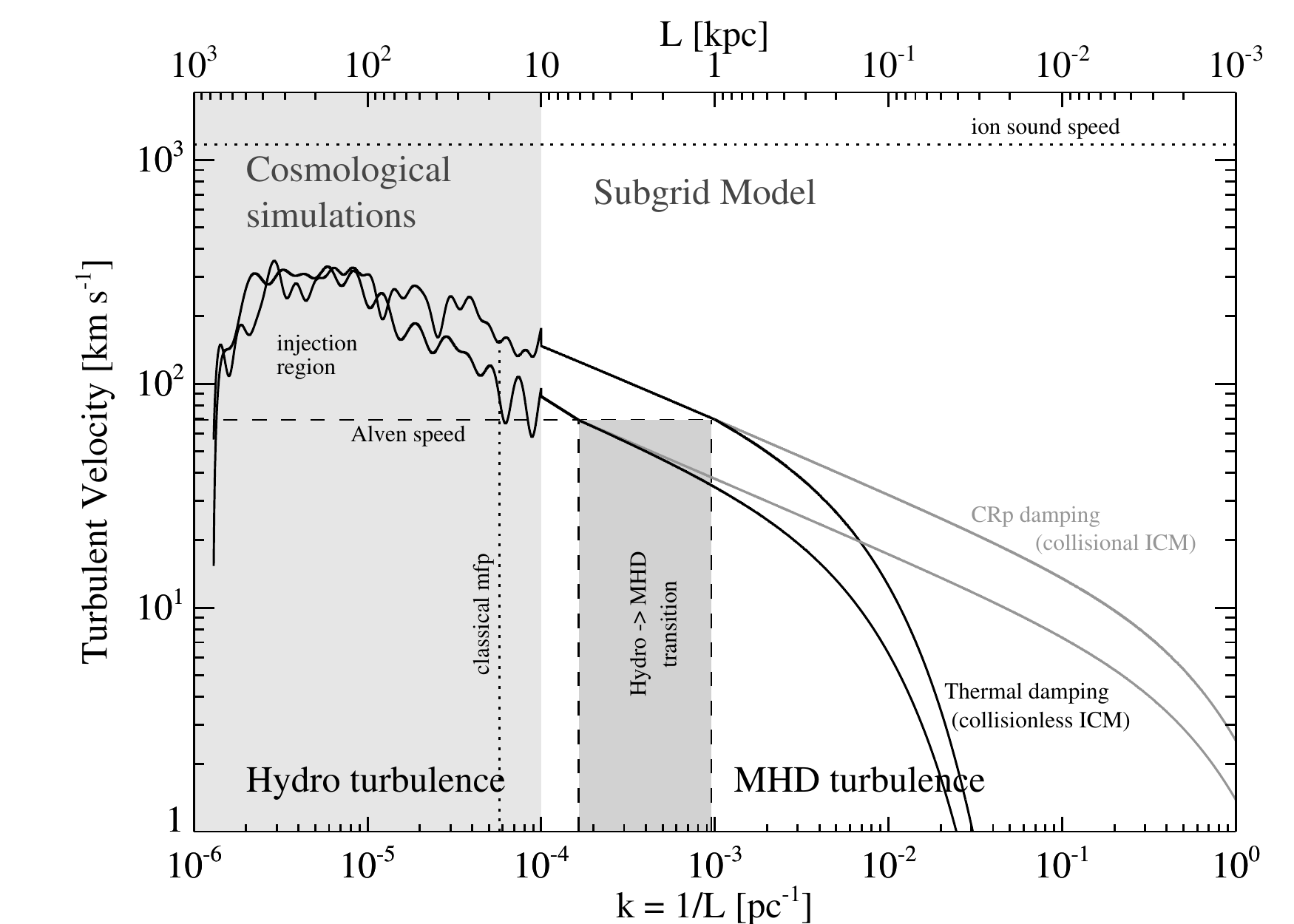}
    \caption{Schematic of the simple turbulence model with standard cluster parameters : $B=1~\mu G$, $n=10^{-3}$~cm, $T=10^{8}$~K, and $\beta_{plasma} \approx 250$. The classical mean free path $\approx200$~kpc, sound speed $c_{s} = 1200$~km s$^{-1}$ and Alfv\'{e}n speed $\approx 100$~km s$^{-1}$ are shown. The Kraichnan turbulence scaling below the simulation scale $\approx 100-10$~kpc is assumed here. In the case of collisional damping, the turbulence cascade damped later, hence the spectrum extends towards smaller scale (grey line) and CR-electrons re-acceleration becomes more efficient. The Figure is taken from \protect\cite{Donnert_2014MNRAS}}
    \label{fig:Schematic_Turbulence_model}
\end{figure}

Mergers of two or more galaxy clusters and motion of sub-halo(s) in cluster/halo deeply stir the ICM, i.e., from core sloshing and shear instabilities. The signatures of the ICM turbulence are largely observed in many numerical simulations \citep{Maier_2009ApJ, Schmidt_2009A&A, Ryu_2010ASPC, Iapichino_2010AIPC, Paul_2011ApJ, Iapichino_2011MmSAI, Vazza_2011A&A, Vazza_2012A&A, Vazza_2017MNRAS, Federrath_2021NatAs} of galaxy clusters. Although there are a few attempts of quantifying the turbulence in merging clusters using X-ray observations \citep[e.g.][]{Roncarelli_2018A&A}. The turbulence motion in clusters of galaxies is defined by its scale, which is almost comparable to the size of cluster cores. Although turbulence in the ICM is majorly sub-sonic, but is sufficient to translate to magneto-hydrodynamic (MHD) turbulences well below the scale at which the Alfv\'{e}n-velocity equals the turbulent velocity. Studies indicate that these turbulent motions are super-Alfv\'{e}nic, with Alfv\'{e}nic Mach number, $M_{A} \gtrsim 5$ \citep{Brunetti_2007MNRAS}. For the assumed ICM conditions, i.e., $B=1~\mu G$, $n=10^{-3}$~cm, $T=10^{8}$~K, and $\beta_{plasma} \approx 250$, in the schematic diagram of the turbulence energy over wavenumber (Fig.~\ref{fig:Schematic_Turbulence_model})  the transition region (dark grey) from hydro to magneto-hydro turbulence has been shown \citep{Donnert_2014MNRAS}.

In addition, the plasma instabilities in the ICM results in various kind of wave, i.e., slow-mode, fast-mode, Alfv\'{e}n-mode, and a few self-excited modes-- slab and whistler. These modes are widely explored as the particle (re)-acceleration mechanisms to explain the radio halo-like emission. The term (re)-acceleration by the plasma driven turbulence is coined here because, the acceleration of charged particles to relativistic energies, directly from the ICM thermal pool, is quite inefficient, rather re-acceleration of pre-existing high energy particles is more efficient. In these stochastic re-acceleration processes, the particles drain energy from plasma turbulence due to resonant interaction with the MHD waves (excited due to plasma instabilities).

In the last few decades, the Alfv\'{e}n-mode and fast-mode are widely explored for understanding particle acceleration. In incompressible turbulence, below the Alfv\'{e}n scale, the acceleration is driven by the Alfv\'{e}n-mode via gyro-resonance \citep{Yan_2002PhRvL, Fujita_2003ApJ, Brunetti_2004MNRAS}. It efficiently re-accelerated the relativistic protons and the relativistic electrons, but the acceleration of charged particles, directly from the thermal pool, is possible only for the thermal protons. The duration of the non-thermal process in clusters of galaxies is largely limited by the back-reaction of protons on the MHD wave, which suppresses the efficient acceleration of relativistic electrons. Alfv\'{e}n acceleration indicates a temporal correlation between the dynamical timescale of merger events and the acceleration timescale of electrons, which is estimated to be one order of magnitude shorter than the former one \citep{Brunetti_2004MNRAS}. In another scenario, the fast-mode induced in the compressible turbulence driven at large scales in the ICM can re-accelerate the CR-electrons to very high energies ($\approx$ a few $GeV$) via Transit-Time-Damping (TTD) resonance \citep{Yan_2004ApJ, Brunetti_2007MNRAS, Yan_2008ApJ, Brunetti_2011aMNRAS, Brunetti_2011bMNRAS}. It assumes quasi-isotropic turbulence cascade and high beta plasma ($\beta_{plasma}$, ratio between plasma pressure to magnetic pressure). The coupling between the magnetic moment of charged particles and the parallel magnetic field gradients essentially causes the acceleration of charged particles. The damping of the compressible turbulence can be with thermal particles and relativistic particles. Fig.~\ref{fig:Schematic_Turbulence_model} \citep{Donnert_2014MNRAS} illustrates that the damping rate with thermal particles (dark full line) is higher than the relativistic particles (light grey line), hence at small scales the cosmic ray re-acceleration becomes more efficient. In all these cases, the acceleration efficiency is constrained by the damping of the turbulence by the CR particles themselves. This modifies the spectrum of the turbulence, and hence the acceleration efficiency (as back-reaction). However, it is still unclear whether such small-scale waves can efficiently be generated in the ICM or not.

One can understand these mechanisms by studying them in controlled large scale cosmological simulations. In one case, this can be achieved by adopting an analytical or a semi-analytical approach, where the power-law for the particle spectra is estimated to be $n_{e}(E_{e}=p_{e}c) \propto E_{e}^{-\delta}$, with the assumed turbulence spectra and the fixed turbulence energy injection, and subsequently compute the radio synchrotron emission using Eq.~\ref{eq:synchrotron_emission} as post-processing on each grid-cell in the cosmological simulation to extract the synthetic synchrotron brightness map \citep{Fang_2016JCAP,Paul_2018arXiv}. The synchrotron emission at frequency, $\nu$, is given as \citep{Rybicki_1986rpabook,Longair-1994hea..book.....L}
\begin{equation}
I_{\nu} = \frac{\sqrt{3} e^{3} B}{m_{e}c^{2}} \int_{E_{\rm min} }^{E_{\rm max}} \int_{0}^{\frac{\pi}{2}}  dE_e\;\; d\theta \;\; sin^{2}\theta\;\;F\left(\frac{\nu}{\nu_c}\right)\;\;n_e(E_e),
\label{eq:synchrotron_emission}
\end{equation}
where $\theta$ is the pitch angle, $F(x)$ is the synchrotron kernel, $\nu_{c}$ is the critical frequency,
\begin{equation}
\nu_{c} = \frac{3}{2} \gamma^{2} \frac{eB}{m_{e} c} sin\theta,
\label{eq:critical_frequency}
\end{equation}
and $B$ is a tangled magnetic field. Here time-evolution of the spectra is not assumed/studied and estimations are drawn at the frozen epoch of large scale structures in cosmological simulations. On the other hand, significant deviations from the power law for the particle spectra are expected during the time evolution. Hence, numerically solving the Fokker-Planck equation to account for the re-acceleration of every particle in the dense gas in the computational domain is essential for more accurate insight. In this direction, \cite{Donnert_2013MNRAS} for the first time attempted to include the complex CR electron re-acceleration physics into the SPH-(Smooth Particle Hydrodynamical) Simulations. They simulate the cosmological flow with high Reynolds numbers to follow the rise and decay of turbulence in the cluster. The evolution of CR electrons is computed in post-processing with all relevant losses and stochastic re-acceleration processes (TTD damping of compressive turbulence in the ICM) by solving the Fokker-Planck equation for every SPH particle. In key observable of their work, in the different states of the merger evolution, they get the variety of observed radio spectra, i.e., flatter, curved and ultra-steep spectrum halos. Interested readers can also refer \cite{Donnert_2014MNRAS} for more insight in numerical solution to the Fokker-Planck equation \citep[Eq.1 of ][]{Donnert_2014MNRAS} and a sub-grid model for compressible magneto-hydrodynamic (MHD) turbulence in the ICM (Fig.~\ref{fig:Schematic_Turbulence_model}).

In the turbulent re-acceleration mechanism, to estimate well-motivated particle-acceleration coefficients and subsequent turbulence scale in different cosmological simulations, one needs to estimate the local turbulence velocity. One simplistic and physically motivated estimate can be the root-mean-square of the velocity dispersion inside the kernel, i.e., subtracting the bulk flow (the weighted or non-weighted average of velocity components) inside the kernel and the root-mean-square of this quantity signifies the turbulent velocity (see in SPH simulations- \cite{Vazza_2006MNRAS, Donnert_2014MNRAS}, Grid-based simulations \cite{Vazza_2009A&A, Vazza_2011A&A, Paul_2018arXiv}. The power spectrum of the velocity field of the simulated ICM can be estimated by applying the standard FFT algorithm to the data at the highest available resolution \citep{Vazza_2009A&A, Vazza_2010gcop, Federrath_2008ApJ, Valdarnini_2011A&A}, which describes the type of turbulence-induced in the ICM. \cite{Vazza_2012A&A} adopted the multiscale filtering algorithms to identify the turbulent motions in hydrodynamical grid simulations (ENZO and FLASH). Most recently, \cite{Federrath_2021NatAs} performed the world's largest, highest-resolution simulations of turbulence in interstellar gas and molecular clouds. It allows for determining the position and width of the sonic scale, which was found to be very close to the theoretical predictions. For the first time, it enables researchers to draw a strong conclusion on the important transition of turbulence from supersonic to subsonic region. Although the initial aim of the simulation is to understand the impact of the turbulence on star formation in molecular clouds, it's the biggest forward leap in understanding the turbulence at sonic scales (\href{https://www.mso.anu.edu.au/~chfeder/pubs/sonic_scale/Federrath_sonic_scale_lowres.mp4}{3D animation available online}).

\subsection{Secondary electrons as the source of synchrotron emission}

As an alternative to the turbulent re-acceleration as the source of high energy  (GeV) electrons, the “hadronic” or “secondary” model has been proposed to account for the production of diffuse radio emission in clusters of galaxies (see, e.g., \citealt{Pfrommer_2004A&A,Keshet_2010ApJ}. This was conceptualized mainly to explain the Mpc scale size of the radio halos that spans almost the entire cluster volume. Similar to volume filling turbulence, Cosmic-ray protons are believed to fill the entire cluster volume \citep{Ensslin_2011A&A,John_2019MNRAS} and can act as the constant injection source for long sustaining emission. The relativistic protons can be produced by the acceleration of thermal proton population by the shocks associated with cosmological structure formation (such as Fermi first order acceleration mechanism at shock fronts), galactic wind, injection by supernovae and the active galactic nuclei (AGN) (\citealt{Ensslin_1998A&A, Volk_1999APh, Ryu_2003ApJ}, for a review see \citealt{2014IJMPD..2330007B}). High energy electrons (or cosmic-ray electrons) responsible for synchrotron radio emission are the decay products (or the secondary particles), produced during the inelastic collision between these relativistic protons and the thermal ions in the ICM \citep{Dennison_1980ApJ, Schlickeiser_1987A&A, Blasi_1999NuPhS, Dolag_2000A&A}. 

The hadronic model somewhat resolves the difficulty in explaining the size of radio halos ($\sim 1$ Mpc); i.e., the required diffusion time for the relativistic electrons to reach such large distances, which is much larger than their radiative lifetime. Since the radiative losses for protons are very low in comparison with electrons, and therefore the lifetime of CR protons is much longer than that of CR electrons. It makes these relativistic protons diffuse to very large scales and hence can accumulate in the cluster environment for a long time. The basic idea of the hadronic model is the continuous injection of cosmic-ray (CR) electrons in the ICM, happening through secondary processes that can  successfully explain the sizes of radio halos.

To numerically model the radio synchrotron emission from the hadronic model: first, the power-law spectrum of the CR protons is considered and thereafter the spectrum of the secondary electrons  is estimated from the proton-proton collisions \citep{Brunetti_2005MNRAS, Pfrommer_2004MNRAS, Donnert_2010aMNRAS, Donnert_2010bMNRAS}.  \citet{Donnert_2010aMNRAS} simulate a scenario similar to the Coma cluster, and numerically predict the radio synchrotron emission from such a cluster, assuming the hadronic model, They report that the purely hadronic model disfavours the current radio-halo observations of the Coma cluster, and moreover, the hadronic model cannot explain the observed high frequency spectral steepening of the radio halo in the Coma cluster. Furthermore, the hadronic model is also incapable of explaining the observed general properties of giant radio halos, such as the observed radio bi-modality in the radio-X-ray correlation plot \citep{Cassano_2013ApJ}, and the overall spectral properties of the halos, which can be very uniform to reasonably patchy \citep{Donnert_2010aMNRAS}. A great deal of debate on this issue can be found in a series of publications, as summarized in \citet{2014IJMPD..2330007B}.

The neutral pions generated during the proton-proton collisions are expected to decay as $\gamma-$ray emission \citep{Blasi_2007IJMPA}, which can be considered as an additional by-product of the hadronic model. It is the $\gamma-$ray emission which gives an additional property to well constraint the secondary model, and the initially considered normalization of CR protons with power-law spectra. Although with the existing $\gamma-$ray telescopes (FERMI telescope), no conclusive evidence of $\gamma-$ray emission yet has been observed in clusters, except for the more recent work by \cite{Baghmanyan_2022MNRAS} showing some indications of extended emission from the nearby Coma cluster. There is a trade-off in the hadronic model, between the radio halo luminosity and the corresponding $\gamma-$ray emissivity if halo emission is purely the cause of secondary particles (CR electrons). Another recent study towards the Coma cluster by \citet{2021A&A...648A..60A} reports the detection of gamma-rays, but they find that the hadronic model is insufficient to explain the radio halo.

\begin{figure*}
\centerline{\includegraphics[width=6.0in]{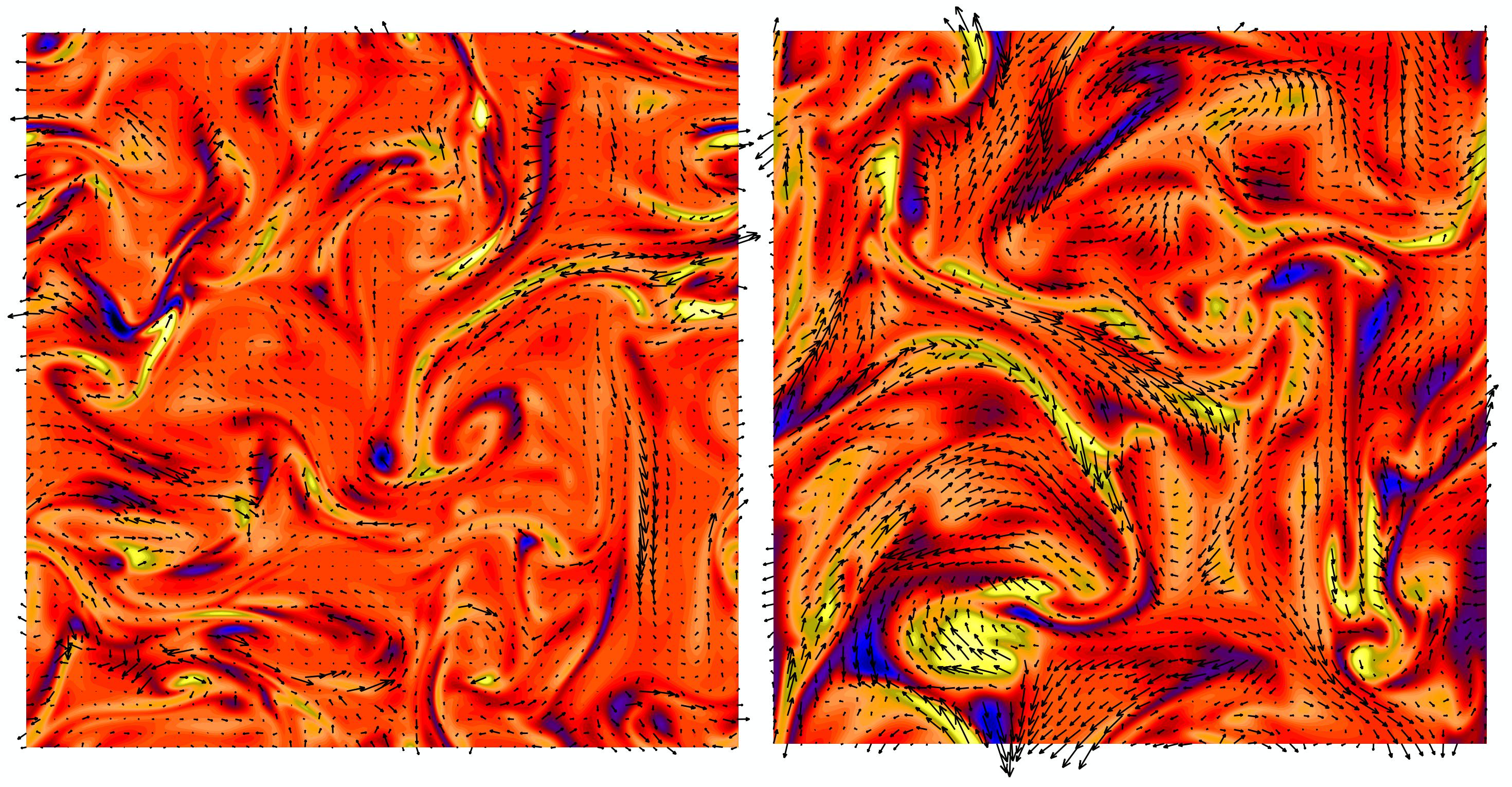}
}
\caption{Two-dimensional snapshots capturing the evolution of 
the magnetic field from initial (left) to the steady state configuration (right), due to 
{\it Fluctuation dynamo} action from a run with $\mathcal{M} = 0.18, \Pm = 1$ 
and turbulence driven on $256\kpc$ scales. Here, the $z$-component of the field is shown in colours in the $x-y$ plane. The fields are strong in blue and yellow regions and negligible in orange regions. The field in the plane of the slices is 
shown with arrows whose length is proportional to the field strength. See \citet{Sur19} and \citet{SBS21} 
for details on the simulation setup and initial conditions.
}\label{flucdyn}
\end{figure*}

Although the numerical predictions for secondary electron models for radio halo emission do not favour the current observations of radio halos, it is found to be an effective model to explain the mini-halo radio structures \citep[e.g.][]{Pfrommer_2004A&A, Ignesti_2020A&A}. \cite{Ignesti_2020A&A} consider the hadronic scenario where they assume the CR-proton injection from the central AGN and CR-proton energy independent diffusion coefficient, to estimate the CR-proton spectrum to finally arrive at the radio emission profile. In their work, the final electron spectrum is computed assuming stationary conditions \citep{Brunetti_2017MNRAS}, which on the first-order approximation can be justified by the cooling time. On this timescale, the CR-protons can diffuse to scales similar to that of mini-halos, within the duty cycles of the AGN. Unfortunately, due to the high complexity of physical processes in the cooling flow clusters, no explicit attempts have been made to numerically simulate such structures.

\subsection{Numerical simulations of magnetic fields in the ICM}\label{sec:mag_field_pol}

Existing studies discussed in Section.~\ref{sec:pol_icm} that compare observations of the Faraday RM and depolarization of cluster radio sources with those derived from numerical models reveal that the magnetic fields in the ICM to be ordered on $\kpc$-scales and have strengths of few $\mkG$. 
Magnetic fields in the ICM can be amplified and maintained at
equipartition\footnote{Here `equipartition' refers to the energy density of
magnetic fields and kinetic energy of gas.} levels via several processes
operating on a large range of scales. For example, on several hundreds of kpc scales, shearing motion and sloshing of gas due to off-axis cluster mergers can drive turbulence \citep{Subramanian_2006MNRAS}.
This turbulence can be driven by the cascade of vortical motions 
produced in oblique shocks as an aftermath of cluster formation and/or 
major collisions \citep{NB99,Subramanian_2006MNRAS, RKCD08, Xu+12, Miniati15, Marinacci+18}. On the other hand, on a few tens of kpc scales, AGN and star-formation 
driven winds in cluster galaxies can drive turbulence \citep{donnert2009, 
dubois2012, pakmor2016, wiene17}. 

In the absence of noticeable rotation, \textit{fluctuation dynamos} are 
ideally suited to amplify dynamically insignificant seed magnetic fields by randomly stretching them in turbulent eddies \citep{Scheko+04, HBD04, Fed+11a, PJR15, SBS18, SF21}. Fluctuation dynamos amplify
the fields to $\mkG$ levels and order them on kpc-scales in a fast and 
efficient way \citep{K68,ZRS90}. The amplification of fields 
by this mechanism occurs on the \textit{eddy-turnover} timescale ($\sim10^8$\,yr), which 
does not require any rotation or density stratification and only relies 
on the random or turbulent nature of the fluid flow. Fig.~\ref{flucdyn} shows that the resulting structure of the fluctuation dynamo generated magnetic fields from numerical simulations of \citet{SBS21} is intermittent and arranged in folds \citep[also see][]{Seta+20}, and are clearly in contrast to Gaussian field distributions \citep[e.g.,][]{Murgia+04}.

\begin{figure*}
\centering
\includegraphics[width=7in]{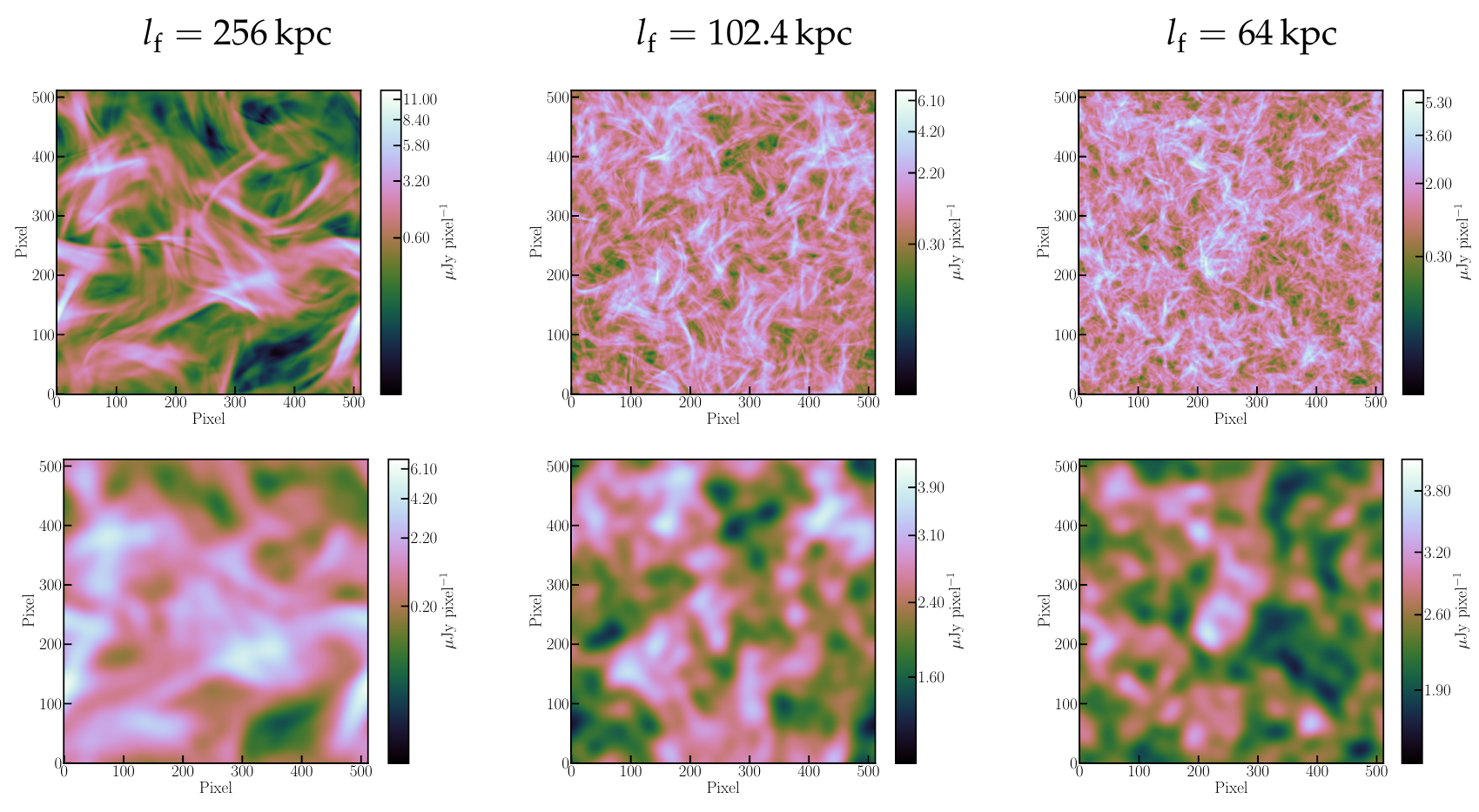}
\caption{\label{fig:turb_driving_effect} Effect of turbulent driving-- the Top row shows 
2D maps of the surface brightness of the total synchrotron emission at the 
native resolution of the simulations. The Bottom row shows the same maps when 
smoothed on $30\kpc$ scales. These figures are from \citet{SBS21}
and \citet{BS21}.}
\end{figure*}

Cosmological simulations of tens of Mpc scales have shed meaningful 
insights into the role of magnetic fields in structure formation in the Universe \citep[see e.g.,][]{RKCD08, donnert2009, Miniati15,Vazza_2017MNRAS, Vazza+18, Marinacci+18, Vazza+21}, and
in understanding the radial distribution of gas and magnetic fields. 
Some of these simulations also confirm the possibility of fluctuation 
dynamo action in galaxy clusters \citep[e.g.,][]{Miniati15,Vazza+18, Marinacci+18, Dominguez+2019}. Although these simulations are a powerful tool and are complementary 
to periodic box, local dynamo simulations, the limited resolution 
achievable to resolve the turbulent scales together with uncontrollable 
numerical dissipation (due to use of ideal MHD) may substantially affect
the evolution and properties of magnetic fields. In fact, \citet{Vazza+18} 
are able to achieve a resolution $\approx 3.95\,{\rm kpc}$ at best by 
using the adaptive mesh refinement technique. Moreover, cosmological 
simulations of galaxy cluster assembly and evolution, at present, are only 
able to follow the dynamical behaviour for only a few tens of dynamical 
times (due to exorbitant computational costs). This implies that the 
magnetic energy would be grossly underestimated, with its value depending 
on the initial seed field and the effective Reynolds number of the simulation
\citep{Beresnyak_2016ApJ}. 
Until computational power and numerical implementation are sufficient
for performing cosmological simulations by aptly capturing the fluctuation
dynamo action as well as resolving scales closer to the dissipation scale, 
numerical simulations of smaller volumes ($< 1\,\rm Mpc^3$) with high 
resolutions, $\mathcal{O}(1\kpc)$, can be useful to gain meaning insights 
on the properties of magnetic fields and their observational signatures \citep[e.g.,][]{Subramanian_2006MNRAS,BS13,Sur19}.

\subsubsection{Polarized emission from fluctuation dynamo action:}
In order to study the expected properties of the polarized synchrotron emission in the ICM due to the action of fluctuation dynamo, high resolution ($1\kpc$)
simulations of a $512^3\kpc^3$ volume, representing the realistic conditions in
the core regions of ICM, have been recently analysed by \citet{SBS21} and
\citet{BS21} using the \texttt{COSMIC} polarization transfer package \citep{Basu+2019}.
By including the effects of frequency-dependent Faraday\footnote{A frequency-dependent reduction in polarization wherein emission
originating from different locations along the line of sight undergo different
degrees of Faraday rotation in the thermal component of the ICM gas.}, and frequency-independent beam depolarization\footnote{A reduction in the polarization caused by cancellation due to turbulent magnetic fields on scales smaller than the telescope beam.}, synthetic 2-D maps of the polarized synchrotron emission in the 0.5--7\,GHz range and maps of Faraday depth were investigated for various representative scales of turbulent driving ($l_f$). In the presence of subsonic turbulence in the ICM, the power spectrum of the Faraday depth map is directly related to the magnetic integral scale ($L_{{\rm int},M}$) defined as,
\begin{equation}
\label{eq:int_scale}
L_{{\rm int},M} = \frac{2\uppi\int [M(k)/k]\,dk}{\int M(k)\,dk}.
\end{equation}
Here, $M(k)$ is the power spectrum of the magnetic energy and $k$ is the wavenumber. Thus, the power spectrum of the map of Faraday depth can provide insights into the driving scale as it is linearly related to $L_{{\rm int},M}$ \citep{BS21}. However, obtaining Faraday depth maps for ICM is tricky due to the faintness of the polarized emission and limitations in analysis techniques, e.g., RM-synthesis \citep[see e.g.,][]{Basu+2019, SBS21}.

Figure~\ref{fig:turb_driving_effect} shows the maps of the total synchrotron intensity for different $l_f$ from \citet{BS21}. The top panels are at the native 1\,kpc resolution of the simulations, and the bottom panels are after smoothing by a telescope beam with FWHM of 30\,kpc corresponding to an angular resolution of 10\,arcsec at redshift $\sim0.2$. It is clear that the filamentary structures on small scales are mostly smeared out in the presence of a telescope beam.
These studies further revealed
that at $\nu \lesssim 1.5\ghz$, a combination of low sensitivity and observation noise results in polarized
emission from bright filamentary structures, wherever detectable, to be confined as clumps that could originate either due to shock compression or from Faraday
depolarization. 

Moreover, at low frequencies ($\nu \lesssim 1.5\ghz$), telescope beam
drastically affects the properties of polarized emission in the presence of
Faraday rotation (shown in the bottom panels of
Figure~\ref{fig:turb_driving_effect}).  Thus, such detection may not provide
adequate information on the global properties of turbulent magnetic fields in
the ICM. On the other hand, the effects of Faraday depolarization and beam smoothing are mitigated for $\nu
\gtrsim 4\ghz$ with polarization remaining largely unaffected and the mean fractional polarization ($\langle p\rangle$) follows the $\langle p\rangle \propto l_f^{1/2}$ relation as expected from random walk of the polarization vector when averaged over a volume \citep{BS21}.

This indicates that observational estimates of, or constrain on, $\langle p\rangle$ at $\nu \gtrsim 4\ghz$ could be effectively used as an
indicator of the driving scale of turbulence in the ICM, strongly underlining the need for high-frequency ($\gtrsim 4\ghz$) polarization
observations of radio halos using current or future radio telescopes. As $\langle p\rangle$ depends on $l_f$ above $\sim4\ghz$, for observations performed with a spatial resolution better than $\sim30\kpc$ (i.e., angular resolution of $\sim10$\,arcsec for a cluster at $z\sim0.2$), $l_f\lesssim20\kpc$ gives rise to $\langle p \rangle\lesssim 0.05$ indicating turbulence driven by gas accretion and star formation feedback from galaxies, $\langle p \rangle$ in the range $0.05\textrm{--}0.2$ would arise for $l_f$ in the range 20--100\,kpc driven by AGN feedback, while for $l_f\gtrsim150\kpc$, $\langle p\rangle \gtrsim 0.2$ is expected, indicating turbulence being driven via vortical motions from cluster mergers. Therefore, to
effectively probe the properties of polarized emission in the ICM and distinguish the different drivers of turbulence, sensitivity to $\langle p\rangle \gtrsim 0.05$ for total intensity surface brightness $\sim2\textrm{--}6\,\rm \upmu Jy\,beam^{-1}$ at $\sim6\textrm{--}10$\,arcsec angular resolution is necessary \citep[see][]{BS21}. Future observations with the SKA in Band\,5a, covering the 4.6 to 8.5\,GHz range, will play a crucial role in the quest of directly unravelling the magnetic field strength and structure in the ICM through polarized emission.

\subsection{Modelling Mechanical heating by AGN}

There is a pool of growing evidence, from both observations and simulations, that heating by the central AGN plays a role in the evolution of galaxy clusters and hence in the formation of the large-scale structural formation \citep{Nath2002,Roychowdhury2005,Guo2008,Gaspari2011,Chaudhuri2012,Chaudhuri2013,Prasad2015,Iqbal2017b}.  This idea gets support from the fact that about 70\% of galaxy clusters host central AGN. The ultra-steep spectrum radio bubbles (AGN remnants) in X-ray detected cavities found in the ICM is a direct proof of AGN energy being injected into ICM. Moreover, the mechanical power of the radio jets is found to be much larger than the synchrotron power \citep{Cavagnolo2010,Godfrey2013}, implying even weak radio sources could produce enough mechanical power to significantly affect the ICM properties up to large radii and prevent gas from cooling. The feedback physics is driven by non-thermal electrons, directly or indirectly, and hence diffuse radio emissions in clusters are correlated to feedback mechanisms.

\begin{figure*}
	\centering
	\begin{minipage}{8.0cm}
		\includegraphics[width = 8.0cm]{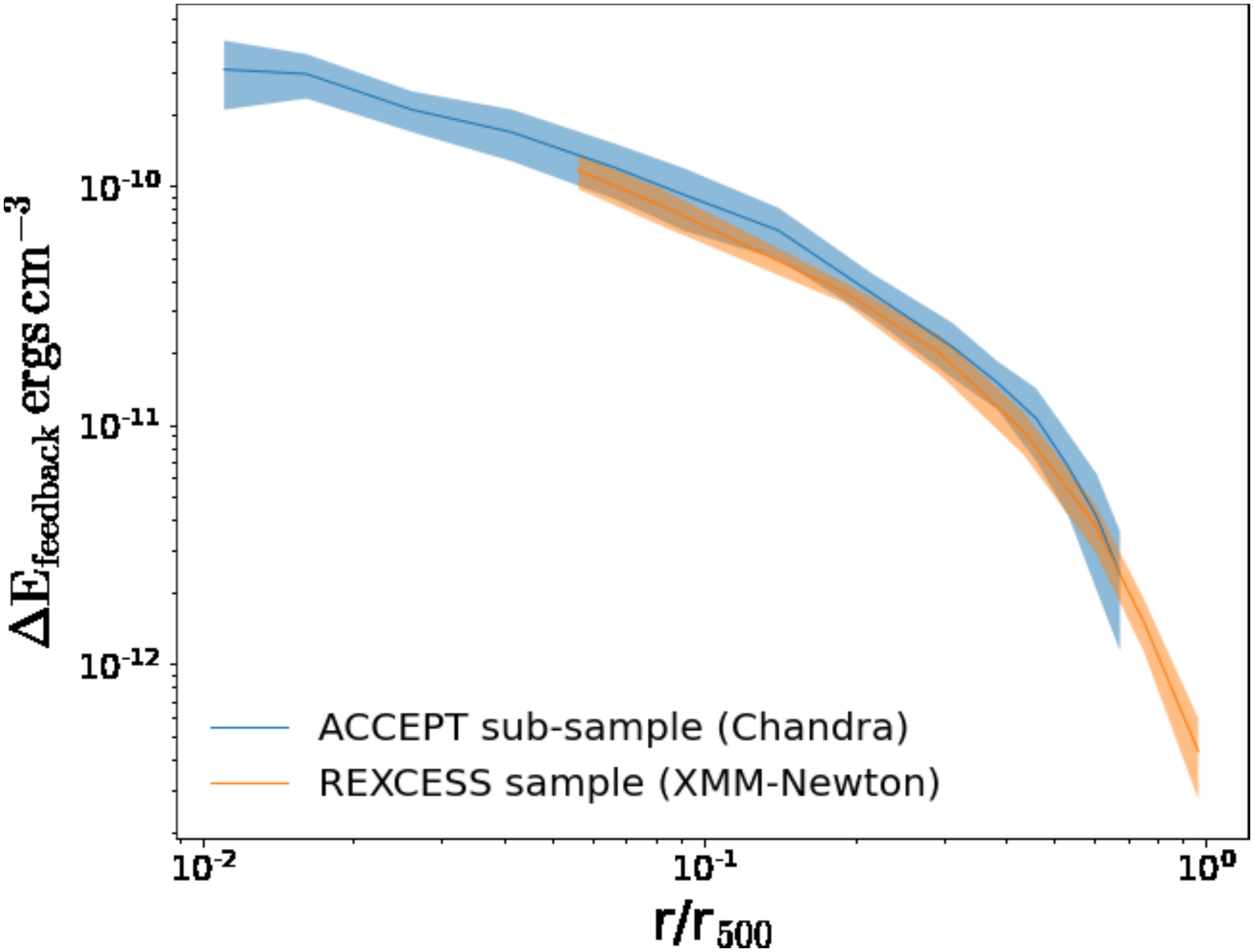}
	\end{minipage}
	\begin{minipage}{8.0cm}
		\includegraphics[width =8.0cm]{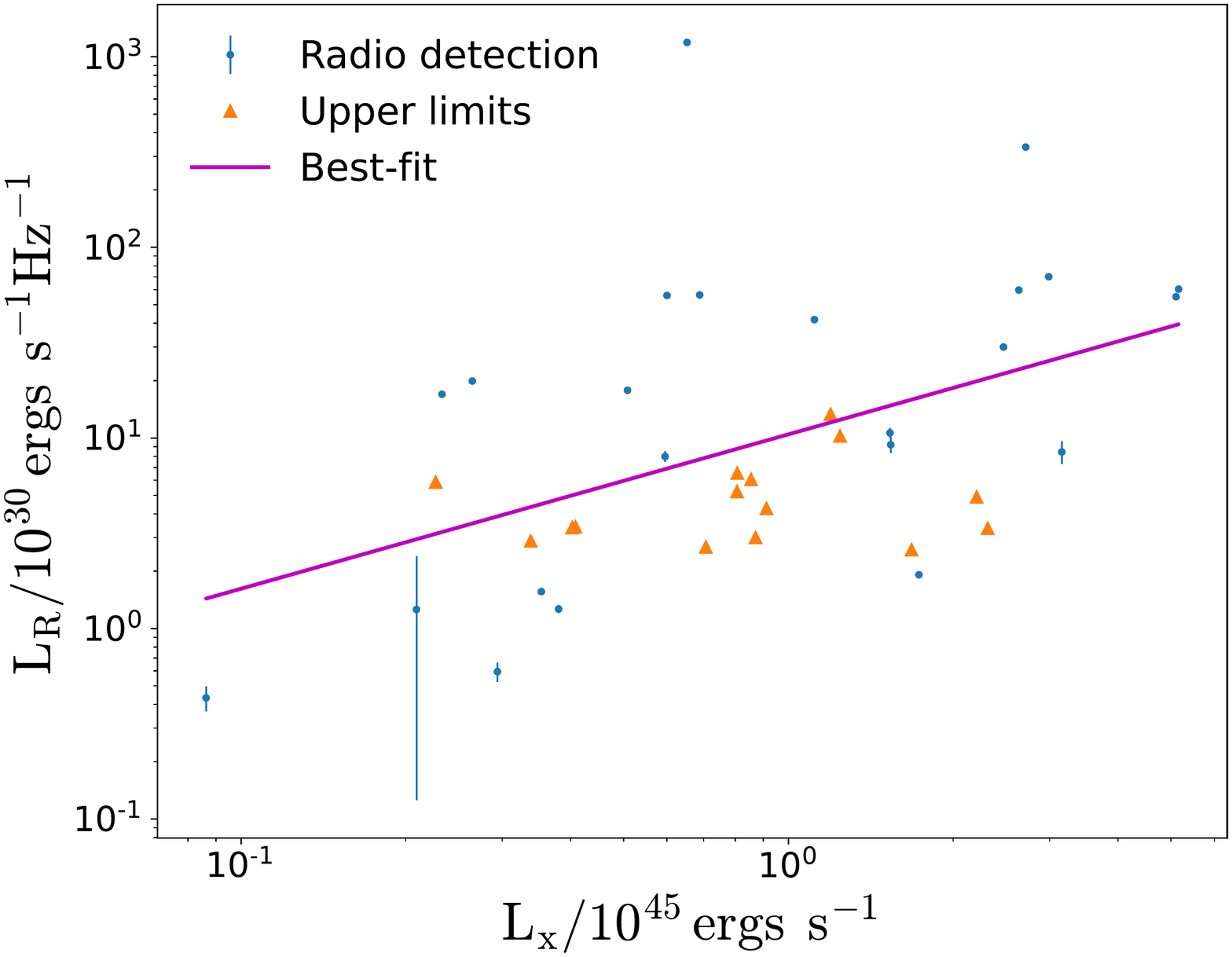}
	\end{minipage}   
	\caption{Left panel:Feedback energy per unit volume  in the ACCEPT sub-sample \citep{Iqbal2018}  and the REXCESS sample \citep{Chaudhuri2013}. Right panel: Correlation between BCG- radio luminosity and X-ray luminosity  in the ACCEPT sub-sample \citep{Iqbal2018}.}
	\label{fig1}
\end{figure*}

By comparing the observed entropy profiles with theoretically expected entropy profiles, based on non-radiative simulations, \cite{Chaudhuri2012,Chaudhuri2013,Iqbal2017} were able to estimate the energy deposition profiles in the ICM. \cite{Chaudhuri2013} found an excess mean energy per particle to be around $\sim 1.6-2.7$ keV up to $r_{500}$.  Using the NRAO/VLA Sky Survey source catalogue, \cite{Chaudhuri2013} found that the total radio luminosity at 1.4~GHz of the central source(s) correlates with feedback energy, albeit with different normalization for cool-core and non-cool-core clusters. \cite{Iqbal2017,Iqbal2017b} estimated energy deposition profiles in galaxy clusters up to the cluster virial radius.  Their results conclusively rule out any pre-heating of the ICM, but agreed with the earlier findings of \cite{Chaudhuri2013}.  In another work, \cite{Iqbal2018} using  Chandra  X-ray and VLA/GMRT radio data in the cluster inner regions found a significant correlation between the BCG radio luminosity ($L_{\rm R}$) and cluster bulk properties. In the left panel in Fig.~\ref{fig1}, we show the median non-gravitational feedback energy as a function of radius for the ACCEPT \cite{Cavagnolo2009} sub-sample and the REXCESS \cite {Rexcess2007,Pratt2010} sample, as estimated in \cite{Iqbal2018}. The profiles are centrally peaked and decrease with radius. The right panel, in Fig.~\ref{fig1}, shows the correlation between the BCG radio luminosity and X-ray luminosity as estimated in \cite{Iqbal2018}. These results suggest that AGN play a dominant role in heating the ICM and that also AGN feedback is intricately linked with cluster radio emission.

There are currently three main AGN feedback mechanisms (models) which can be explored without resorting to complex hydro-dynamical simulations: (i) the acoustic model \citep{Fabian2003,Fabian2005}, (ii) the effervescent model \citep{Ruszkowski2002,Roychowdhury2004}  and (iii) the cosmic-ray model \citep{Guo2008,Yutaka2011}.  In the acoustic model, the dissipation of energy into the ICM is due to sound waves induced by the central AGN activity while in the effervescent model, the AGN injects non-thermal electrons filled buoyant bubbles into the ICM which then heat the ambient medium by doing $pdV$ work.  The cosmic-ray model assumes that cosmic rays are injected by the AGN,  which can then heat the cluster core by amplifying Alfv\'{e}n waves \citep{Guo2008,Yutaka2011}. Although all the models assume the central AGN to be the main source of feedback energy, they are radically different with regard to the physics of dissipation of energy into the ICM. All the three models, in principle, can be constrained by studying the mechanical jet power and radio luminosity ($L_{\rm jet}-L_{\rm R}$) scaling relations \citep{Cavagnolo2010,Godfrey2013} and the thermal properties of the ICM  using X-rays and radio measurements. Unfortunately, with current X-ray and radio observations, it is not possible to differentiate unambiguously between these models \citep{Iqbal_2022arXiv}.  However, upcoming high-quality data from large surveys with uGMRT \& SKA  (in radio), CMB-S4 (in mm) and Athena (in X-rays)  will provide us new insights on AGN feedback of the ICM up to $z \sim 2$.

\section{Updated taxonomy of cluster radio sources}\label{sec:taxonomy}

As discussed earlier in section~\ref{sec:current_class}, historically the cluster radio sources were classified broadly into radio halos, relics and mini-halos mostly based on their physical characteristics \citep{Ensslin_1998A&A,Giovannini_1999NewA,Feretti_2008LNP,Feretti_2012A&ARv}. However, with the improved sensitivity and wide spectral coverage of telescopes in recent years, a variety of extended low-brightness features have been commonly detected  in clusters, slowly blurring the lines between radio galaxies
and diffuse cluster radio sources. Many of these sources are morphologically similar to well known cluster radio sources but sharply differing in their spectral properties, especially in steepness and curvature. This has necessitated the inclusion of revived AGN fossil plasma sources, phoenices, and GReET together as a new class of cluster radio sources, making the overall classification scheme a very complex one \citep{Weeren_2019SSRv}, a mixture of multiple parameters such as location, morphology, spectra etc.

Nevertheless, it is evident that, where radio halos, mini-halos and peripheral relics are in general detected with a power-law spectrum, AGN relics or fossil plasma sources, phoenices etc. exhibit a curved spectrum. As per our current understanding and the suggested theoretical models (see Sec.~\ref{sec:theoretical_models_simulations}), while, the power-law spectrum favours an in-situ particle acceleration mechanism, spectral curvature due to the ageing of electrons indicates the dying or revival of synchrotron emitting electrons. Therefore, the major factors that determine the energy distribution and the appearance, are the initial electron energy spectrum (steepness) and the dynamics of the cluster medium that drives the particle (re-)acceleration engines. 

Since a substantial number of cluster radio sources are now available with their spectral information in wider frequency bands and the list is only expected to grow in future, we propose here a novel idea for classifying these sources based on their origin, specifically on the type of source (injection) electrons and cluster dynamics. This would in future help us to rationalize the classification and the nomenclature of a large variety of diffuse radio sources being observed. In this proposed taxonomy, we call a type of source as (i) Active cluster radio source if the sources show a power-law spectrum with no obvious spectral curvature. Since the power-law spectrum is the characteristics of freshly accelerated non-thermal particles, it indicates the presence of in-situ and active particle acceleration engines in these sources, such as usual radio-halos, relics and mini-halos. The sources will be called as the (ii) Dying cluster radio sources, if they are of ultra-steep spectrum nature with spectral curvature and age $\lesssim100$~Myr with no obvious particle acceleration engine found in the system. AGN relics, ultra-steep radio halos with spectral curvature etc. would fall in this group. And a third one is defined as the (iii) Revived cluster radio sources, in case the emission comes from the revival of fossil non-thermal electrons, where cluster shock or turbulence is involved. However, they act as the revival engine rather than an accelerating engine and keep the curved spectrum nature of the injected fossil or aged electrons, intact. The basic feature of these sources are an ultra-steep spectrum, spectral curvature and a spectral age of usually beyond $100$~Myr, such as radio phoenices, revived ultra-steep sources by turbulence, revived buoyant radio bubbles etc. 

While the existing classification scheme would remain relevant for immediate categorization of the overall cluster radio sources, our proposed taxonomy would help in classifying them further into more fundamental level. Such a classification would not be host-specific, rather would be applicable to a wider variety of radio emissions in various large scale structures. However, it should be noted that the classification always evolve in time with the availability of more data and as our understanding of the systems progressively improves.

\section{On the role of uGMRT and expectations with the SKA}\label{sec:role_uGMRT_SKA}

For a complete understanding of the physics of the ICM and the interplay between the galaxies and the ICM, it is going to be important to make deep continuum observations, in particular focussing on polarization and spectral line studies, of these components across a wide range of frequency bands. Recent galaxy cluster surveys such as LoTSS at 144 MHz and MGCLS at 1283 MHz (Sec.~\ref{sec:surveys}) have resulted in numerous discoveries of diffuse radio emission from galaxy clusters. The uGMRT, given its full  coverage of the LOFAR sky, in addition to a significant overlap with the MeerKAT sky, is well suited for providing follow-up observations, in the frequency range of 150--1400 MHz, for obtaining spectra of such diffuse radio sources. Polarization studies at sub-GHz frequencies are challenging, but will be of interest to study sources that have ultra-steep spectra. The uGMRT bands 4 and 3 are being tested to make polarization measurements \citep{2021MNRAS.507..991S} and will be of interest to study magnetic fields in sources such as radio relics and phoenices.

The $\upmu$Jy level sensitivity, that the SKA will achieve, will transform our understanding of the statistical properties of large scale structures, like galaxy clusters down to $10^{14}$\(\textup{M}_\odot\) and beyond z $\geq$1. This  has been limited until now due to the sensitivity and resolution constraints imposed by current generation instruments \citep{Cassano_2015}. It will also help us to probe the distant Universe via cluster gravitational lensing, particularly at radio wavelengths where the number of known radio-emitting lensed galaxies (a few mJy, up to $\upmu$Jy level) are scarce from z$\geq$1 (\citealt{2015aska.confE..84M}, Sec.~2.1.8 by Pandey-Pommier et al. in \citealt{Acero_2017arXiv} and Pandey-Pommier et al. SKA Users case, pvt. communication).

Deep continuum imaging of galaxies belonging to clusters is also necessary to study the effect of ICM on galaxies and vice versa. SKA continuum surveys will provide unprecedented sensitivities to study cluster fields in depth. Continuum imaging at 0.2–0.6 arcsec, and 3 $\upmu$Jy/beam sensitivity on clusters along with spectral line (HI) surveys of lensed galaxies at SKA LOW (0.05--0.35 GHz), SKA1-MID Band 1 (0.35-1.05) $\&$ 2 (0.95-1.76) frequencies will not only help to discover the intra-cluster filaments at lower redshifts z $\le$0.1 and investigate the matter distribution within clusters up to z $\sim$1, but also may discover a new population of background star-forming galaxies up to (0 $\le$ z $\le$6) via imaging. Further, for higher redshift lensed galaxies, (z $\le$5) continuum imaging and spectral line observations, in Band 5 (4.6 -15.3~GHz) with 0.05--0.1 arcsec and 0.3$\upmu$Jy level sensitivity, will provide insights into the cold molecular gas reservoirs (CO, HCN, HCO+, etc.) \citep{2013ARA&A..51..105C} that trace the sites of ongoing star formation. Additionally, the deep HI and continuum imaging provided with the uGMRT in Band 5 (1.4 GHz) with 2.5 $\upmu$Jy/beam sensitivity and 2.3 arcsec resolution will help us trace the impact of the extreme cluster environment on the atomic gas (HI) reservoirs, and also disentangle the HI emission from the cluster central regions and the filaments (Pandey-Pommier et al. SKA Users case, pvt. communication).

\section{Summary and Conclusions}\label{sec:summary}
Radio observations of galaxy clusters are direct probes of a variety of phenomena on scales ranging from galaxies to the large scale structures. Diffuse radio sources found in galaxy clusters represent signatures of synchrotron emission delineating the magnetic fields in a cluster-scale system, along with the presence of relativistic electrons in the ICM. These sources have low surface brightness ($\sim\mu\rm{Jy~arcsec^{-2}}$ at GHz), steep spectra ($\alpha < -1$ with $S_\nu \propto \nu^{\alpha}$), and extend up to several tens of arcminutes (in low redshift sources). 
Thus low frequency telescopes with good sensitivity to extended structures, with low rms noise, are essential for these studies. 

In this work we have reviewed the current understanding of this field using current radio observations, along with theoretical models and simulations, and outline the priorities for upcoming and already operational radio telescopes. We summarize as follows:
\begin{enumerate}
    \item The detailed scientific study of physical phenomena
    in galaxy groups and clusters require  radio observations  with high resolution (reaching sub-kpc scales) and with good sensitivity to extended structures (few arcminutes to a degree). Thus, large telescopes such as the VLA, WSRT and GMRT have played an important role in the past several decades in these studies. The Upgraded GMRT, LOFAR and MeerKAT have recently become operational, and are providing new insights on the physics of the galaxy clusters and beyond. The SKA is a next generation radio telescope which will further revolutionize the field.
    \item Radio surveys targeting galaxy clusters in the past two decades have thrown light on the statistics of occurrence of diffuse radio emission in clusters. Currently, several all-sky surveys, such as those with the LOFAR, are regularly generating numerous new discoveries that need detailed follow-up to uncover their nature and origin.
    \item Diffuse radio emission from clusters of galaxies is broadly classified into radio halos, relics and mini-halos. There are other diffuse filamentary extended radio sources found in galaxy clusters that result from activities or mergers of radio galaxies. The interaction between the cluster ICM and the radio sources can lead to radio sources being revived. Radio phoenices, turbulence-revived sources and buoyant radio bubbles are sources in this category. 
    \item The focus of radio surveys is shifting to low mass clusters and groups where the state of the magnetic field and extent of the generation of cosmic rays is poorly known. 
    \item At the other extreme of spatial extent, superclusters of galaxies which can stretch beyond 10-100~Mpc belong to a scale where the origin and nature of magnetic fields are poorly known. The detection of synchrotron-emitting radio bridges between pairs of clusters indicate that substantial magnetic fields do exist on such scales, and the dense environment of superclusters may host large scale radio sources that can provide the direct detection of such magnetic fields. Currently, both superclusters and binary clusters are being surveyed in the radio bands.
    \item Theoretical models have proposed mechanisms that explain the origin and nature of diffuse radio sources to a large extent, though there are unsolved challenges. The Diffusive Shock Acceleration model falls short of explaining the observed radio relics unless the seed relativistic electron population is included. Several variants of the turbulent reacceleration model, such as using compressible and incompressible turbulence, Alfv\'{e}n modes and fast modes, have been proposed for explaining radio halos. Efforts are ongoing to numerically solve the Fokker-Planck equation to follow the reacceleration of every particle in a typical simulation box. However, overcoming numerical limitations and obtaining predictions specific to a variety of merger scenarios will be needed.
    \item Recent numerical simulations of magnetic fields show intermittent morphological signatures, arranged in folds, in contrast to Gaussian field distributions that are typically considered in the literature. Cosmological simulations are reaching unprecedented resolution scales (current down to a few kpc), though more needs to be done to capture the fluctuation dynamo and resolution scales of the order of dissipation scales. Simulations of observations show that polarization observations at 4~GHz and above could potentially be used to find the driving scales of turbulence.
    \item By studying the radio sources in the central galaxies in clusters and groups, constraints can be obtained on the mechanical heating of the cluster cores by AGN. This will lead to the understanding of the physics of the interaction of the thermal and non-thermal plasma in the ICM, and the accretion processes associated with the  supermassive black holes at the cores of the brightest cluster and group galaxies.
    \item The Upgraded GMRT with its wide frequency coverage will play a major role as a follow-up instrument for the newly discovered radio sources with LOFAR and MeerKAT, paving the way for SKA.
    \item With the advent of SKA, significant progress in understanding the open issues regarding the origin and nature of magnetic fields and cosmic rays on the scales of clusters and beyond is expected.
\end{enumerate}

\appendix

\section*{Acknowledgements}
S.~Paul wants to thank the DST INSPIRE Faculty Scheme (code: IF-12/PH-44) for funding his research group. R.~Kale acknowledges the support of the Department of Atomic Energy, Government of India, under project no. 12-R\&D-TFR-5.02-0700.
S.~Sur acknowledges computing time awarded the CDAC National Param supercomputing 
facility, India, under the grant `Hydromagnetic-Turbulence-PR' and the use of 
the High Performance Computing (HPC) resources made available by the Computer 
Centre of the Indian Institute of Astrophysics. S.~Sur further thanks the 
Science and Engineering Research Board (SERB) of the Department of Science \& 
Technology (DST), Government of India, for support through research grant ECR/2017/001535. A.\,Basu acknowledges the computing facility and support provided at the Th\"uringer Landessternwarte, Tautenburg, Germany. The work of A. Iqbal was supported by CNES, France. MP acknowledges the support of CEFIPRA foundation under the project 6504-3. MR acknowledges financial support from Ministry of Science and Technology of Taiwan (MOST 109-2112-M-007-037-MY3).

\vspace{-1em}

\bibliographystyle{mnras}
\bibliography{jaa_cluster,ruta-test,msbib,AGN_bib,references_vp,models,sr-bib}

\end{document}